\documentclass[10pt,a4paper]{article}
 \pdfoutput=1
\usepackage[latin1]{inputenc}
\usepackage{amsmath}
\usepackage{amsfonts}
\usepackage{amssymb}
\usepackage{graphicx}
\usepackage[usenames,dvipsnames,svgnames,table]{xcolor}
\usepackage{subfig}
\usepackage{natbib}
\usepackage{tikz}
\usepackage{url}
\usepackage{multirow}

\definecolor{gray1}{rgb}{1,0,0}
\definecolor{gray2}{rgb}{1,.5,0}
\definecolor{gray3}{rgb}{1,1,0}
\definecolor{gray4}{rgb}{0,1,0}
\definecolor{gray5}{rgb}{0,0,1}

\newcommand{\1}{\mathbf{1}} 
\newcommand{\E}{\mathbb{E}} 
\newcommand{\iidsim}{\overset{iid}{\sim}}

\newcommand{\N}{\text{N}} 
\newcommand{\Gam}{\text{Gam}} 
\newcommand{\DP}{\text{DP}} 
\newcommand{\bc}{\mathbf{c}} 
\newcommand{\bC}{\mathbf{C}} 
\newcommand{\bchat}{\widehat{\mathbf{c}}}
\newcommand{\VI}{\text{VI}} 
\newcommand{\B}{\text{B}} 
\newcommand{\tB}{\tilde{\text{B}}} 
\newcommand{\R}{\text{R}} 
\newcommand{\En}{\text{H}} 
\newcommand{\I}{\text{I}} 
\newcommand{\0}{\mathbf{0}} 

\newcommand{\argmin}{\operatornamewithlimits{argmin}}

\newtheorem{theorem}{Theorem}[section]
\newtheorem{property}[theorem]{Property}
\newtheorem{definition}[theorem]{Definition}

\newcommand{\qed}{\nobreak \ifvmode \relax \else
      \ifdim\lastskip<1.5em \hskip-\lastskip
      \hskip1.5em plus0em minus0.5em \fi \nobreak
      \vrule height0.75em width0.5em depth0.25em\fi}
      
      \urldef\wade\url{https://www2.warwick.ac.uk/fac/sci/statistics/staff/academic-research/wade/}

\title{
Bayesian cluster analysis: Point estimation and credible balls}
\author{Sara Wade and Zoubin Ghahramani}
\begin{document}
\maketitle

\begin{abstract}
Clustering is widely studied in statistics and machine learning, with applications in a variety of fields. As opposed to popular algorithms such as agglomerative hierarchical clustering or k-means which return a single clustering solution, Bayesian nonparametric models provide a posterior over the entire space of partitions, allowing one to assess statistical properties, such as uncertainty on the number of clusters. However, an important problem is how to summarize the posterior; the huge dimension of partition space and difficulties in visualizing it add to this problem. In a Bayesian analysis, the posterior of a real-valued parameter of interest is often summarized by reporting a point estimate such as the posterior mean along with 95\% credible intervals to characterize uncertainty. In this paper, we extend these ideas to develop appropriate point estimates and credible sets to summarize the posterior of the clustering structure based on decision and information theoretic techniques.
\end{abstract}


\section{Introduction}\label{sec:Intro}
Clustering is widely studied in statistics and machine learning, with applications in a variety of fields. Numerous models and algorithms for clustering exist, and new studies which apply these methods to cluster new datasets or develop novel models or algorithms are constantly being produced. Classical algorithms such as agglomerative hierarchical clustering or the k-means algorithm (\cite{Hartigan79}) are popular but only explore a nested subset of partitions or require specifying the number of clusters apriori. Moreover, they are largely heuristic and not based on formal models, prohibiting the use of statistical tools, for example, in determining the number of clusters.

Model-based clustering methods utilize finite mixture models, where each mixture component corresponds to a cluster (\cite{FR02}). Problems of determining the number of clusters and the component probability distribution can be dealt with through statistical model selection, for example, through various information criteria. The expectation-maximization (EM) algorithm is typically used for maximum likelihood estimation (MLE) of the mixture model parameters. Given the MLEs of the parameters, the posterior probability that a data point belongs to a class can be computed through Bayes rule. The cluster assignment of the data point corresponds to the class with maximal posterior probability, with the corresponding posterior probability reported as a measure of uncertainty. Importantly, however, this measure of uncertainty ignores uncertainty in the parameter estimates. As opposed to MLE, Bayesian mixture models incorporate prior information on the parameters and allow one to assess uncertainty in the clustering structure unconditional on the parameter estimates.


Bayesian nonparametric mixture models assume that the number of components is infinite. As opposed to finite mixture models, this not only avoids specification of the number of components but also allows the number of clusters present in the data to grow unboundedly as more data is collected. Bayesian nonparametric mixture models induce a random partition model (\cite{Q1}) of the data points into clusters, and the posterior of the random partition reflects our belief and uncertainty of the clustering structure given the data. 


However, an important problem in Bayesian nonparametric cluster analysis is how to summarize this posterior; indeed, often the first question one asks is what is an appropriate point estimate of the clustering structure based on the posterior. Such a point estimate is useful for concisely representing the posterior and often needed in applications. Moreover, a characterization of the uncertainty around this point estimate would be desirable in many applications. Even in studies of Bayesian nonparametric models where the latent partition is used simply as a tool to construct flexible models, such as in mixture models for density estimation (\cite{Lo}), it is important to understand the behavior of the latent partition to improve understanding of the model. To do so, the researcher needs to be equipped with appropriate summary tools for the posterior of the partition. 

Inference in Bayesian nonparametric partition models usually relies on Markov chain Monte Carlo (MCMC) techniques, which produce a large number of partitions that represent approximate samples from the posterior. Due to the huge dimension of the partition space and the fact that many of these partitions are quite similar differing only in a few data points, the posterior is typically spread out across a large number of partitions. Clearly, describing all the unique partitions sampled would be unfeasible, further emphasizing the need for appropriate summary tools to communicate our findings.

In a typical Bayesian analysis, the posterior of a univariate parameter of interest is often summarized by reporting a point estimate such as the posterior mean, median, or mode, along with a 95\% credible interval to characterize uncertainty. In this paper, we aim to extend these ideas to develop summary tools for the posterior on partitions. In particular, we seek to answer the two questions: 1) What is an appropriate point estimate of the partition based on the posterior? 2) Can we construct a 95\% credible region around this point estimate to characterize our uncertainty?

We first focus on the problem of finding an appropriate point estimate. A simple solution is to use the posterior mode. If the marginal likelihood of the data given the partition, that is with all mixture component parameters integrated out, and the prior of the partition are available in closed form, the posterior mode can be estimated based on the MCMC output by the sampled partition which maximizes the non-normalized posterior. In practice, a closed form for the marginal likelihood or prior is often unavailable, specifically, if conjugate priors for the component specific parameters do not exist or are not utilized or hyperpriors are assigned to any hyperparameters. More generally, the posterior mode can be found by reporting the partition visited most frequently in the sampler. Yet this approach can be problematic, as producing reliable frequency counts is intractable due to the huge dimension of the partition space. In fact,  in many examples, the MCMC chain does not visit a partition more than once. To overcome this, alternative search techniques have been developed to locate the posterior mode (\cite{Heller05}, \cite{Heard06}, \cite{Dahl09}, \cite{RBL14}). 
However, it is well-known that the mode can be unrepresentative of the center of a distribution.

Alternative methods have been proposed based on the posterior similarity matrix. For a sample size of $N$, the elements of this $N$ by $N$ matrix represent the probability that two data points are in the same cluster, which can be estimated by the proportion of MCMC samples that cluster the two data points together. Then, classical hierarchical or partitioning algorithms are applied based on the similarity matrix 
(\cite{Med02}, \cite{Med04}, \cite{RDGW09}, \cite{Molitor10}). These methods have the disadvantage of being ad-hoc. 

A more elegant solution is based on decision theory. In this case, one defines a loss function over clusterings. 
The optimal point estimate is that which minimizes the posterior expectation of the loss function. For example, for a real-valued parameter $\theta$, the optimal point estimate is the posterior mean under the squared error loss $L_2(\theta, \widehat{\theta})=(\theta-\widehat{\theta})^2$ and the posterior median under the absolute error loss $L_1(\theta, \widehat{\theta})=|\theta-\widehat{\theta}|$. 

The question to answer then becomes what is an appropriate loss function on the space of clusterings. The 0-1 loss function, a simple choice which leads to the posterior mode as the point estimate, is not ideal as it does not take into account the similarity between two clusterings. More general loss functions were developed by \cite{Binder78}, and the so-called Binder's loss, which measures the disagreements in all possible pairs of observations between the true and estimated clusterings, was studied in a Bayesian nonparametric setting by \cite{Lau07}. Alternative loss functions considered in Bayesian nonparametrics can be found in \cite{Quintana03} and \cite{Fritsch09}.

In this paper, we propose to use the variation of information developed by \cite{Meila07} as a loss function in a Bayesian nonparametric setting. Both the variation of information and Binder's loss possess the desirable properties of being metrics on the space of partitions and being \textit{aligned} with the lattice of partitions. We provide a detailed comparison of these two metrics and discuss the advantages of the variation of information over Binder's loss as a loss function in Bayesian cluster analysis. Additionally, we propose a novel algorithm to locate the optimal partition, taking advantage of the fact that both metrics are aligned on the space of partitions.

Next, to address the problem of characterizing uncertainty around the point estimate, we propose to construct a credible ball around the point estimate. As both Binder's loss and the variation of information are metrics on the partition space, we can easily construct such a ball. Interestingly, the two metrics can produce very different credible balls, and we discuss this in detail. In existing literature, quantifications of uncertainty include reporting a heat map of the estimated posterior similarity matrix. However, there is no precise quantification of how much uncertainty is represented by the posterior similarity matrix, and in a comparison with the 95\% credible balls, we find that the uncertainty is under-represented by the posterior similarity matrix. Finally, we provide an algorithm to construct the credible ball and discuss ways to depict or report it.

The paper is organized as follows. Section \ref{sec:review} provides a review of Bayesian nonparametric clustering and existing point estimates of the clustering structure from a decision theoretic approach. In Section \ref{sec:compare_VIvBL}, we give a detailed comparison of two loss functions, Binder's loss and the variation of information, pointing out advantages of the latter. 
The optimal point estimate under the variation of information is derived in Section \ref{sec:pe_VI} and a novel algorithm to locate the optimal partition is proposed. In Section \ref{sec:CB}, we construct a credible ball around the point estimate to characterize posterior uncertainty and discuss how to compute and depict it. Finally, simulated and real examples are provided in Section \ref{sec:examples}.

\section{Review}\label{sec:review}

This section provides a review of Bayesian nonparametric clustering models and existing point estimates of the clustering in literature.

\subsection{Bayesian nonparametric clustering}

Mixture models are one of the most popular modeling tools in Bayesian nonparametrics. The data is assumed conditionally i.i.d. with density
$$ f(y|P) = \int K(y|\theta) dP(\theta),$$
where $K(y|\theta)$ is a specified parametric density on the sample space with mixing parameter $\theta \in \Theta$ and $P$ is a probability measure on $\Theta$. In a Bayesian setting, the model is completed with a prior on the unknown parameter, which in this case, is the unknown mixing measure. In the most general setting, this parameter $P$ can be any probability measure on $\Theta$, requiring a nonparametric prior. 
Typically the nonparametric prior has discrete realizations almost surely (a.s.) with
$$P= \sum_{j=1}^\infty w_j \delta_{\theta_j} \text{ a.s.},$$
where it is often assumed that the weights $(w_j)$ and atoms $(\theta_j)$ are independent and the $\theta_j$ are i.i.d. from some base measure $P_0$. Thus, the density is modeled with a countably infinite mixture model
$$ f(y|P)=\sum_{j=1}^\infty w_j K(y|\theta_j).$$

Since $P$ is discrete a.s., this model induces a latent partitioning $\bc$ of the data where two data points belong to the same cluster if they are generated from the same mixture component. The partition can be represented by $\bc=(C_1,\ldots, C_{k_N})$, where $C_j$ contains the indices of data points in the $j^{\text{th}}$ cluster and $k_N$ is the number of clusters in the sample of size $N$. Alternatively, the partition can be represented by $\bc=(c_1,\ldots,c_N)$, where $c_n=j$ if the $n^{\text{th}}$ data point is in the $j^{\text{th}}$ cluster. 

A key difference with finite mixture models is that the number of mixture components is infinite; this allows the data to determine the number of clusters 
$k_N$ present in the data, which can grow unboundedly with the data. 
Letting  $\mathbf{y}_j= \lbrace y_n \rbrace_{n \in C_j}$, the marginal likelihood for the data $y_{1:N}$ given the partition is
$$ f(y_{1:N}| \bc)=\prod_{j=1}^{k_N} m(\mathbf{y}_j)=\prod_{j=1}^{k_N} \int \prod_{n \in C_j} K(y_n|\theta) dP_0(\theta).$$

The posterior of the partition, which reflects our beliefs and uncertainty in the clustering given the data, is simply proportional to the prior times the marginal likelihood 
\begin{align}
 p( \bc |y_{1:N}) \propto p(\bc) \prod_{j=1}^{k_N} m(\mathbf{y}_j) ,
\label{eq:post_clus}
\end{align}
where the prior of the partition is obtained from the selected prior on the mixing measure. For example, a Dirichlet process prior (\cite{Ferg}) for $P$ with mass parameter $\alpha$ corresponds to
$$ p( \bc)= \frac{\Gamma(\alpha)}{\Gamma(\alpha+N)}\alpha^{k_N} \prod_{j=1}^{k_N} \Gamma(n_j), $$
where $n_j=|C_j|$ is the number of data points in cluster $j$. Various other priors developed in Bayesian nonparametric literature can be considered for the mixing measure $P$, such as the Pitman-Yor process (\cite{PitmanYor}), also known as the two-parameter Poisson-Dirichlet process, or the normalized generalized Gamma process or more generally, a prior within the class of normalized completely random measures, Poisson-Kingman models (\cite{Pitman03}), or stick-breaking priors (\cite{IJ}). See \cite{LP10} for an overview. 

In general, the marginal likelihood of the data given the partition or the prior of the partition used to compute the posterior in (\ref{eq:post_clus}) may not be available in closed form. Moreover, there are
$$S_{N,k}= \frac{1}{k!} \sum_{j=0}^{k} (-1)^j \left( \begin{array}{c} k \\ j \end{array} \right) (k-j)^N,$$
a Stirling number of the second kind, ways to partition the $N$ data points into $k$ groups and
$$B_N= \sum_{k=1}^{N} S_{N,k},$$ a Bell number, possible partitions of the $N$ data points. Even for small $N$, this number is very large, which makes computation of the posterior intractable for the simplest choice of prior and likelihood. Thus, MCMC techniques are typically employed, such as the marginal samplers described by \cite{Neal} with extensions in \cite{FT13} for normalized completely random measures and in \cite{LFT16} for $\sigma$-stable Poisson-Kingman models; the conditional samplers described in \cite{IJ}, \cite{PaR08}, or \cite{KGW11}, with extensions in \cite{FT13} for normalized completely random measures and in \cite{FW12} for $\sigma$-stable Poisson-Kingman models; or the recently introduced class of hybrid samplers for $\sigma$-stable Poisson-Kingman models in \cite{LFT15}. These algorithms produce approximate samples $(\bc^m)_{m=1}^M$ from the posterior (\ref{eq:post_clus}). Clearly, describing all the posterior samples is infeasible, and our aim is to develop appropriate summary tools to characterize the posterior.

Extensions of Bayesian nonparametric mixture models are numerous and allow one to model increasingly complex data. These include extensions for partially exchangeable data (\cite{Teh}), inclusion of covariates (\cite{MacTR}), time dependent data (\cite{GS}), and spatially dependent data (\cite{Duan}) to name a few. See \cite{MuQ04} and \cite{DunBNP} for an overview. These extensions also induce latent clusterings of the observations, and the summary tools developed here are applicable for these settings as well.

\subsection{Point estimation for clustering}

Firstly, we seek a point estimate of the clustering that is representative of the posterior, which may be of direct interest to the researcher or, more generally, important for understanding the behavior of the posterior. From decision theory, a point estimate is obtained by specifying a loss function $L(\bc, \widehat{\bc})$, which measures the loss of estimating the true clustering $\bc$ with $ \widehat{\bc}$. Since the true clustering is unknown, the loss is averaged across all possible true clusterings, where the loss associated to each potential true clustering is weighted by its posterior probability. The point estimate $ \bc^*$ corresponds to the estimate which minimizes the posterior expected loss,
$$ \bc^*=\argmin_{\widehat{\bc}} \E[L(\bc,\widehat{\bc})|y_{1:N}]=\argmin_{\widehat{\bc}} \sum_{\bc} L(\bc,\widehat{\bc})p(\bc | y_{1:N}).$$

A simple choice for the loss function is the 0-1 loss, $L_{0-1}( \bc, \widehat{\bc})=\1(\bc \neq \widehat{\bc})$, which assumes a loss of 0 if the estimate is equal to the truth and a loss of 1 otherwise. Under the 0-1 loss, the optimal point estimate is the posterior mode. 
However, this loss function is unsatisfactory because it doesn't take into account similarity between two clusterings; a partition which differs from the truth in the allocation of only one observation is penalized the same as a partition which differs from the truth in the allocation of many observations. Moreover, it is well-known that the mode can be unrepresentative of the center of a distribution. Thus, more general loss functions are needed.

However, constructing a more general loss is not straightforward because, as pointed out by \cite{Binder78}, the loss function should satisfy basic principles such as invariance to permutations of the data point indices and invariance to permutations of the cluster labels for both the true and estimated clusterings. \citeauthor{Binder78} notes that this first condition implies that the loss is a function of the counts $n_{i\,j}= |C_i \cap \widehat{C}_j |$, which is the cardinality of the intersection between $C_i$, the set of data point indices in cluster $i$ under $\bc$, and $\widehat{C}_j$, the set of data point indices in cluster $j$ under $\widehat{\bc}$, for $i=1,\ldots,k_N$ and $j=1, \ldots, \widehat{k}_N$; the notation $k_N$ and $\widehat{k}_N$ represents the number of clusters in $\bc$ and $\widehat{\bc}$, respectively. 
He explores loss functions satisfying these principles, starting with simple functions of the counts $n_{i\,j}$. The so-called Binder's loss is a quadratic function of the counts, which for all possible pairs of observations, penalizes the two errors of allocating two observations to different clusters when they should be in the same cluster or allocating them to the same cluster when they should be in different clusters:
$$ \B(\bc,\widehat{\bc})= \sum_{n<n'} l_1 \1(c_n=c_{n'})\1(\widehat{c}_n \neq \widehat{c}_{n'})+l_2 \1(c_n \neq c_{n'})\1(\widehat{c}_n = \widehat{c}_{n'}).$$
If the two types of errors are penalized equally, $l_1=l_2=1$, then
$$ \B(\bc,\widehat{\bc})= \frac{1}{2} \left( \sum_{i=1}^{k_N} n_{i\,+}^2+ \sum_{j=1}^{\widehat{k}_N} n_{+\, j}^2 -2 \sum_{i=1}^{k_N}\sum_{j=1}^{\widehat{k}_N} n_{i\,j}^2\right),$$
where $n_{i\,+}=\sum_{j} n_{i\,j}$ and $n_{+\,j}=\sum_{i}n_{i\,j}$.
Under Binder's loss with $l_1=l_2$, the optimal partition $\bc^*$ is the partition $\bc$ which minimizes
\begin{align*}
 \sum_{n<n'} \left\lvert \1(c_n=c_{n'}) - p_{n\,n'} \right\rvert,
 \end{align*}
or equivalently, the partition $\bc$ which minimizes
\begin{align}
\sum_{n<n'} \left( \1(c_n=c_{n'}) - p_{n\,n'} \right)^2, \label{eq:minBinder}
\end{align}
where $p_{n\,n'}=P(c_n=c_{n'}|y_{1:N})$ is the posterior probability that two observations $n$ and $n'$ are clustered together. 
This loss function was first studied in Bayesian nonparametrics by \cite{Lau07}. We note that in earlier work \cite{Dahl06} considered minimization of (\ref{eq:minBinder}) but without the connection to Binder's loss and the decision theoretic approach.

Binder's loss counts the total number of disagreements, $D$, in the  ${ N \choose 2}$ possible pairs of observations. The Rand index (\cite{Rand71}), a cluster comparison criterion, is defined as the number of agreements, $A$, in all possible pairs divided by the total number of possible pairs. Since $D+A= { N \choose 2}$, Binder's loss and the Rand index, denoted $\R(\bc, \widehat{\bc})$, are related:
$$ \B(\bc,\widehat{\bc})= (1-\R(\bc, \widehat{\bc})){\footnotesize { N \choose 2}},$$
and the point estimate obtained from minimizing the posterior expected Binder's loss is equivalent to the point estimate obtained from maximizing the posterior expected Rand's index. Motivated by this connection, \cite{Fritsch09} consider maximizing the adjusted Rand index, introduced by \cite{HA85} to correct the Rand index for chance. An alternative loss function is explored by \cite{Quintana03} specifically for the problem of outlier detection.

\section{A comparison of the variation of information and Binder's loss}\label{sec:compare_VIvBL}

\cite{Meila07} 
introduces the \textit{variation of information} (VI) for cluster comparison, which is constructed from information theory and compares the information in two clusterings with the information shared between the two clusterings. More formally, the VI is defined as
\begin{align*}
&\VI(\bc,\widehat{\bc})=\En(\bc)+\En(\widehat{\bc})-2\I(\bc,\widehat{\bc})\\
&=-\sum_{i=1}^{k_N} \frac{n_{i\,+}}{N}\log\left(\frac{n_{i\,+}}{N}\right)-\sum_{j=1}^{\widehat{k}_N} \frac{n_{+\,j}}{N}\log\left(\frac{n_{+\,j}}{N}\right)-2\sum_{i=1}^{k_N}\sum_{j=1}^{\widehat{k}_N} \frac{n_{i\,j}}{N}\log\left(\frac{n_{i\,j}N}{n_{i\,+}n_{+\,j}}\right),
\end{align*}
where $\log$ denotes $\log$ base $2$. 
The first two terms represent the entropy of the two clusterings, which measures the uncertainty in bits of the cluster allocation of an unknown randomly chosen data point given a particular clustering of the data points. The last term is the mutual information between the two clusterings and measures the reduction in the uncertainty of the cluster allocation of a data point in $\bc$ when we are told its cluster allocation in $\widehat{\bc}$. The VI ranges from 0 to $\log(N)$. A review of extensions of the VI to normalize or correct for chance are discussed in \cite{VEB10}. However, some desirable properties of the VI are lost under these extensions.

In this paper, we propose to use the VI as a loss function. Note that since 
$\I(\bc,\widehat{\bc})=\En(\bc)+\En(\widehat{\bc})-\En(\bc,\widehat{\bc}),$
we can write
\begin{align*}
\VI(\bc,\widehat{\bc})&=\En(\bc)+\En(\widehat{\bc})-2\En(\bc)-2\En(\widehat{\bc})+2\En(\bc,\widehat{\bc}),\\
&=-\En(\bc)-\En(\widehat{\bc})+2\En(\bc,\widehat{\bc}),\\
&=\sum_{i=1}^{k_N} \frac{n_{i\,+}}{N}\log\left(\frac{n_{i\,+}}{N}\right)+\sum_{j=1}^{\widehat{k}_N} \frac{n_{+\,j}}{N}\log\left(\frac{n_{+\,j}}{N}\right)-2\sum_{i=1}^{k_N}\sum_{j=1}^{\widehat{k}_N} \frac{n_{i\,j}}{N}\log\left(\frac{n_{i\,j}}{N}\right).
\end{align*}
We provide a detailed comparison with an $N$-invariant version of Binder's loss, defined as
$$ \tB(\bc,\widehat{\bc})= \frac{2}{N^2}\B(\bc,\widehat{\bc})=  \sum_{i=1}^{k_N} \left(\frac{n_{i\,+}}{N}\right)^2+ \sum_{j=1}^{\widehat{k}_N} \left(\frac{n_{+\, j}}{N}\right)^2 -2 \sum_{i=1}^{k_N}\sum_{j=1}^{\widehat{k}_N} \left(\frac{n_{i\,j}}{N}\right)^2.$$
Both loss functions are considered $N$-invariant as they only depend on $N$ through the proportions $n_{i\,j}/N$. 
We focus on these two loss functions as they satisfy several desirable properties.

The first important property is that both VI and $ \tB$ are metrics on the space of partitions.
\begin{property}
Both VI and $\tB$ are metrics on the space of partitions.
\end{property}
\noindent A proof for VI can be found in \cite{Meila07}. For $\tB$, the proof results from the fact that $\tB$ can be derived as the Hamming distance between the binary representation of the clusterings.

\begin{figure}
\centering
\begin{tikzpicture}[scale=.7]
  \tikzstyle{every node}=[draw=none]
\draw (0,0) node (n1) {$\lbrace1,2,3,4\rbrace$} ;
 \draw (-7.5,-1.5) node (n2){$\lbrace1 \rbrace \lbrace 2,3,4\rbrace$};
\draw (-5,-1.5) node (n3){$\lbrace2 \rbrace \lbrace 1,3,4\rbrace$};
\draw (-2.5,-1.5) node (n4){$\lbrace3 \rbrace \lbrace 1,2,4\rbrace$};
\draw (0,-1.5) node (n5){$\lbrace4 \rbrace \lbrace 1,2,3\rbrace$};
\draw (2.5,-1.5) node (n6){$\lbrace1,2 \rbrace \lbrace 3,4\rbrace$};
\draw (5,-1.5) node (n7){$\lbrace1,3 \rbrace \lbrace 2,4\rbrace$};
\draw (7.5,-1.5) node (n8){$\lbrace1,4 \rbrace \lbrace 2,3\rbrace$};
 \draw (-7.5,-3) node (n9){$\lbrace 1 \rbrace \lbrace 2 \rbrace \lbrace3,4\rbrace$};
\draw (-4.5,-3) node (n10){$\lbrace 1 \rbrace \lbrace 3 \rbrace \lbrace 2,4\rbrace$};
\draw (-1.5,-3) node (n11){$\lbrace 1 \rbrace \lbrace 4 \rbrace \lbrace 2,3\rbrace$};
\draw (1.5,-3) node (n12){$\lbrace 2 \rbrace \lbrace 3 \rbrace \lbrace 1,4\rbrace$};
\draw (4.5,-3) node (n13){$\lbrace 2 \rbrace \lbrace 4 \rbrace \lbrace 1,3\rbrace$};
\draw (7.5,-3) node (n14){$\lbrace 3 \rbrace \lbrace 4 \rbrace \lbrace 1,2\rbrace$};
 \draw (0,-4.5) node (n15){$\lbrace 1 \rbrace \lbrace 2 \rbrace \lbrace3 \rbrace \lbrace 4\rbrace$};

  \foreach \from/\to in {n1/n2,n1/n3,n1/n4,n1/n5,n1/n6,n1/n7,n1/n8,n2/n9,n2/n10,n2/n11,n3/n9,n3/n12,n3/n13,n4/n10,n4/n12,n4/n14,n5/n11,n5/n13,n5/n14,n6/n9,n6/n14,n7/n10,n7/n13,n8/n11,n8/n12,n9/n15,n10/n15,n11/n15,n12/n15,n13/n15,n14/n15}
    \draw (\from) -- (\to);

\end{tikzpicture}
\caption{Hasse diagram for the lattice of partitions with a sample of size $N=4$. A line is drawn from $\bc$ up to $\widehat{\bc}$ when $\bc$ is covered by $\widehat{\bc}$.}
\label{fig:hasse_N4}
\end{figure}
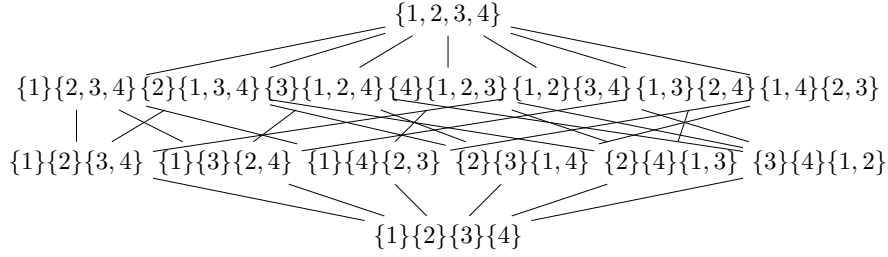

The next properties involve first viewing the space of partitions as a partially ordered set. In particular, consider the space of partitions $\bC$ and the binary relation $\leq$ on $\bC$ defined by set containment, i.e. for $ \bc, \bchat \in \bC$, $ \bc \leq \bchat$ if for all $i=1,\ldots, k_N$, $C_i \subseteq \widehat{C}_j$ for some $j \in \lbrace 1,\ldots, \widehat{k}_N \rbrace$. The partition space $\bC$ equipped with $\leq$ is a partially ordered set. 

For any $\bc, \bchat \in \bC$, $\bc$ is \textit{covered} by $\bchat$, denoted $\bc \prec \bchat$, if $\bc < \bchat$ and there is no $ \widehat{\bchat} \in \bC$ such that $ \bc < \widehat{\bchat} < \bchat$. This covering relation is used to define the \textit{Hasse diagram}, where the elements of $\bC$ are represented as nodes of a graph and a line is drawn from $\bc$ up to $\bchat$ when $\bc \prec \bchat$. An example of the Hasse diagram for $N=4$ is depicted in Figure \ref{fig:hasse_N4}.

The space of partitions possesses an even richer structure; it forms a lattice. This follows from the fact that every pair of partitions has a \textit{greatest lower bound} and \textit{least upper bound}; for a subset $ \mathbf{S} \subseteq \bC$, an element $\bc \in \bC$ is an upper bound for $ \mathbf{S}$ if $ \mathbf{s}\leq \mathbf{c}$ for all $\mathbf{s} \in \mathbf{S}$, and $\bc \in \bC$ is the least upper bound for $ \mathbf{S}$, denoted $\bc=\text{l.u.b.}(\mathbf{S})$, if $\bc$ is an upper bound for $\mathbf{S}$ and $ \bc \leq \mathbf{c}'$ for all upper bounds $\mathbf{c}'$ of $\mathbf{S}$. A lower bound and the greatest lower bound  for a subset $ \mathbf{S} \subseteq \bC$ are similarly defined, the latter denoted by $\text{g.l.b.}(\mathbf{S})$.
We define the operators $\wedge$, called the meet, and $\vee$, called the join, as $\bc \wedge \bchat=\text{g.l.b.}(\bc,\bchat)$ and  $\bc \vee \bchat=\text{l.u.b.}(\bc,\bchat)$. Following the conventions of lattice theory, we will use $ \mathbf{1}$ to denote the greatest element of the lattice of partitions, i.e. the partition with every observation in one cluster $\bc=( \lbrace 1,\ldots,N \rbrace)$, and  $\mathbf{0}$ to denote the least element of the lattice of partitions, i.e. the partition with every observation in its own cluster $\bc= (\lbrace 1\rbrace,\ldots, \lbrace N \rbrace)$. See \cite{Nation91} for more details on lattice theory and the Supplementary Material for specific details on the lattice of partitions.

A desirable property is that both VI and $\tB$ are \textit{aligned} with the lattice of partitions. Specifically, both metrics are \textit{vertically aligned} in the Hasse diagram; if $\widehat{\bchat}$ is connected up to $ \bchat$ and $ \bchat$ is connected up to $\bc$, then the distance between $\widehat{\bchat}$ and $\bc$ is the vertical sum of the distances between $\widehat{\bchat}$ and $\bchat$ and between $\bchat$ and $\bc$ (see Property \ref{prop:VertAlign}). And, both metrics are \textit{horizontally aligned}; the distance between any two partitions is the horizontal sum of the distances between each partition and the meet of the two partitions (see Property \ref{prop:HorzAlign}).

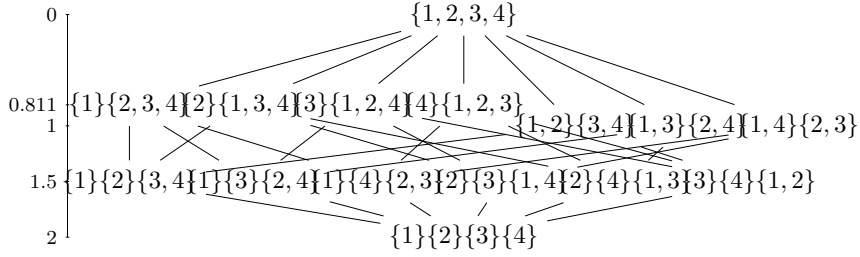
\begin{figure}[!h]
\centering
\begin{tikzpicture}[scale=.7]
  \tikzstyle{every node}=[draw=none]
\draw (8,0) node (n1) {$\lbrace1,2,3,4\rbrace$} ;
 \draw (1.25,-1.825) node (n2){$\lbrace1 \rbrace \lbrace 2,3,4\rbrace$};
\draw (3.5,-1.825) node (n3){$\lbrace2 \rbrace \lbrace 1,3,4\rbrace$};
\draw (5.75,-1.825) node (n4){$\lbrace3 \rbrace \lbrace 1,2,4\rbrace$};
\draw (8,-1.825) node (n5){$\lbrace4 \rbrace \lbrace 1,2,3\rbrace$};
\draw (10.25,-2.25) node (n6){$\lbrace1,2 \rbrace \lbrace 3,4\rbrace$};
\draw (12.5,-2.25) node (n7){$\lbrace1,3 \rbrace \lbrace 2,4\rbrace$};
\draw (14.75,-2.25) node (n8){$\lbrace1,4 \rbrace \lbrace 2,3\rbrace$};
 \draw (1.25,-3.375) node (n9){$\lbrace 1 \rbrace \lbrace 2 \rbrace \lbrace3,4\rbrace$};
\draw (3.75,-3.375) node (n10){$\lbrace 1 \rbrace \lbrace 3 \rbrace \lbrace 2,4\rbrace$};
\draw (6.25,-3.375) node (n11){$\lbrace 1 \rbrace \lbrace 4 \rbrace \lbrace 2,3\rbrace$};
\draw (8.75,-3.375) node (n12){$\lbrace 2 \rbrace \lbrace 3 \rbrace \lbrace 1,4\rbrace$};
\draw (11.25,-3.375) node (n13){$\lbrace 2 \rbrace \lbrace 4 \rbrace \lbrace 1,3\rbrace$};
\draw (13.75,-3.375) node (n14){$\lbrace 3 \rbrace \lbrace 4 \rbrace \lbrace 1,2\rbrace$};
 \draw (8,-4.5) node (n15){$\lbrace 1 \rbrace \lbrace 2 \rbrace \lbrace3 \rbrace \lbrace 4\rbrace$};

  \foreach \from/\to in {n1/n2,n1/n3,n1/n4,n1/n5,n1/n6,n1/n7,n1/n8,n2/n9,n2/n10,n2/n11,n3/n9,n3/n12,n3/n13,n4/n10,n4/n12,n4/n14,n5/n11,n5/n13,n5/n14,n6/n9,n6/n14,n7/n10,n7/n13,n8/n11,n8/n12,n9/n15,n10/n15,n11/n15,n12/n15,n13/n15,n14/n15}
    \draw (\from) -- (\to);
    
\draw (0,0) -- (0,-4.5);
\draw  (1pt,0 cm) -- (-1pt,0 cm) node[anchor=east] {{\footnotesize $0$}};
\draw  (1pt,-1.825 cm) -- (-1pt,-1.825 cm) node[anchor=east] {{\footnotesize $0.811$}};
\draw  (1pt,-2.25 cm) -- (-1pt,-2.25 cm) node[anchor=east] {{\footnotesize $1$}};
\draw  (1pt,-3.375 cm) -- (-1pt,-3.375 cm) node[anchor=east] {{\footnotesize $1.5$}};
\draw  (1pt,-4.5 cm) -- (-1pt,-4.5 cm) node[anchor=east] {{\footnotesize $2$}};

\end{tikzpicture}
\caption{Hasse diagram stretched by VI with a sample of size $N=4$. Note $2-\frac{3}{4}\log(3)\approx 0.811$. From the $\VI$ stretched Hasse diagram, we can determine the distance between any two partitions. Example: if $\bc=(\lbrace 1,2 \rbrace, \lbrace 3,4 \rbrace)$ and $\bchat=(\lbrace 1 \rbrace ,\lbrace 3 \rbrace ,\lbrace 2,4 \rbrace)$, then $\bc \wedge \bchat = (\lbrace 1 \rbrace ,\lbrace 2 \rbrace ,\lbrace 3 \rbrace, \lbrace 4 \rbrace)$ and $d(\bc,\widehat{\bc})=d(\bc \wedge \bchat,\1)-d(\bc,\1)+d(\bc \wedge \bchat,\1)-d(\bchat,\1)=2-1+2-1.5=1.5$.}
\label{fig:hasse_vi_N4}
\end{figure}

\begin{figure}[!h]
\centering
\begin{tikzpicture}[scale=.7]
  \tikzstyle{every node}=[draw=none]
\draw (8,0) node (n1) {$\lbrace1,2,3,4\rbrace$} ;
 \draw (1.25,-2.25) node (n2){$\lbrace1 \rbrace \lbrace 2,3,4\rbrace$};
\draw (3.5,-2.25) node (n3){$\lbrace2 \rbrace \lbrace 1,3,4\rbrace$};
\draw (5.75,-2.25) node (n4){$\lbrace3 \rbrace \lbrace 1,2,4\rbrace$};
\draw (8,-2.25) node (n5){$\lbrace4 \rbrace \lbrace 1,2,3\rbrace$};
\draw (10.25,-3) node (n6){$\lbrace1,2 \rbrace \lbrace 3,4\rbrace$};
\draw (12.5,-3) node (n7){$\lbrace1,3 \rbrace \lbrace 2,4\rbrace$};
\draw (14.75,-3) node (n8){$\lbrace1,4 \rbrace \lbrace 2,3\rbrace$};
 \draw (1.25,-3.75) node (n9){$\lbrace 1 \rbrace \lbrace 2 \rbrace \lbrace3,4\rbrace$};
\draw (3.75,-3.75) node (n10){$\lbrace 1 \rbrace \lbrace 3 \rbrace \lbrace 2,4\rbrace$};
\draw (6.25,-3.75) node (n11){$\lbrace 1 \rbrace \lbrace 4 \rbrace \lbrace 2,3\rbrace$};
\draw (8.75,-3.75) node (n12){$\lbrace 2 \rbrace \lbrace 3 \rbrace \lbrace 1,4\rbrace$};
\draw (11.25,-3.75) node (n13){$\lbrace 2 \rbrace \lbrace 4 \rbrace \lbrace 1,3\rbrace$};
\draw (13.75,-3.75) node (n14){$\lbrace 3 \rbrace \lbrace 4 \rbrace \lbrace 1,2\rbrace$};
 \draw (8,-4.5) node (n15){$\lbrace 1 \rbrace \lbrace 2 \rbrace \lbrace3 \rbrace \lbrace 4\rbrace$};

  \foreach \from/\to in {n1/n2,n1/n3,n1/n4,n1/n5,n1/n6,n1/n7,n1/n8,n2/n9,n2/n10,n2/n11,n3/n9,n3/n12,n3/n13,n4/n10,n4/n12,n4/n14,n5/n11,n5/n13,n5/n14,n6/n9,n6/n14,n7/n10,n7/n13,n8/n11,n8/n12,n9/n15,n10/n15,n11/n15,n12/n15,n13/n15,n14/n15}
    \draw (\from) -- (\to);
    
\draw (0,0) -- (0,-4.5);
\draw  (1pt,0 cm) -- (-1pt,0 cm) node[anchor=east] {{\footnotesize $0$}};
\draw  (1pt,-2.25 cm) -- (-1pt,-2.25 cm) node[anchor=east] {{\footnotesize $0.375$}};
\draw  (1pt,-3 cm) -- (-1pt,-3 cm) node[anchor=east] {{\footnotesize $0.5$}};
\draw  (1pt,-3.75 cm) -- (-1pt,-3.75 cm) node[anchor=east] {{\footnotesize $0.625$}};
\draw  (1pt,-4.5 cm) -- (-1pt,-4.5 cm) node[anchor=east] {{\footnotesize $0.75$}};

\end{tikzpicture}
\caption{Hasse diagram stretched by $\tB$ with a sample of size $N=4$. From the $\tB$ stretched Hasse diagram, we can determine the distance between any two partitions. Example: if $\bc=(\lbrace 1,2 \rbrace, \lbrace 3,4 \rbrace)$ and $\bchat=(\lbrace 1 \rbrace ,\lbrace 3 \rbrace ,\lbrace 2,4 \rbrace)$, then $\bc \wedge \bchat = (\lbrace 1 \rbrace ,\lbrace 2 \rbrace ,\lbrace 3 \rbrace, \lbrace 4 \rbrace)$ and $d(\bc,\widehat{\bc})=d(\bc \wedge \bchat,\1)-d(\bc,\1)+d(\bc \wedge \bchat,\1)-d(\bchat,\1)=0.75-0.5+0.75-0.625=0.375$.}
\label{fig:hasse_bl_N4}
\end{figure}

\begin{property} \label{prop:VertAlign}
For both VI and $\tB$, if $ \bc \geq \bchat \geq \widehat{\bchat}$, 
then
$$ d(\bc, \widehat{\bchat})= d(\bc, \bchat)+ d(\bchat,\widehat{\bchat}).$$
\end{property}
\begin{property} \label{prop:HorzAlign}
For both VI and $\tB$,
$$ d(\bc, \bchat)= d(\bc, \bchat\wedge \bc)+ d(\bchat,\bchat\wedge \bc).$$
\end{property}
\noindent Proofs can be found in the Supplementary Material. These two properties imply that if the Hasse diagram is stretched to reflect the distance between any partition and $\1$, the distance between any two partitions can be easily determined from the \textit{stretched Hasse diagram}. Figures \ref{fig:hasse_vi_N4} and \ref{fig:hasse_bl_N4} depict the Hasse diagram for $N=4$ in Figure \ref{fig:hasse_N4} stretched according to VI and $\tB$ respectively. 

From the stretched Hasse diagram, we gain several insights into the similarities and differences between the two metrics. An evident difference is the scale of the two diagrams.

\begin{property} \label{prop:scale}
A distance on partitions satisfying Properties \ref{prop:VertAlign} and \ref{prop:HorzAlign} has the property that for any two partitions $\bc$ and $\bchat$,
\begin{align*}
d(\bc, \bchat) \leq d(\1,\0).
\end{align*}
Thus,
$$ \VI(\bc, \bchat) \leq \log(N)\quad \text{  and  } \quad \tB(\bc, \bchat) \leq 1-\frac{1}{N}.$$
\end{property}
\noindent A proof can be found in the Supplementary Material. In both cases, the bound on the distance between two clusterings depends on the sample size $N$. However, the behavior of this bound is very different; for VI, it approaches infinity as $N\rightarrow \infty$, and for $\tB$, it approaches one as $N \rightarrow \infty$. As $N$ grows, the number of total partitions $B_N$ increases drastically. Thus, it is sensible that the bound on the metric grows as the size of the space grows. In particular, $\1$ and $\0$ become more distant as $N \rightarrow \infty$, as there is an increasing number, $B_N-2$, of partitions between these two extremes; for $\tB$, the loss of estimating one of these extremes with the other approaches the fixed number one, while for VI, the loss approaches infinity.

From the stretched Hasse diagram in Figures \ref{fig:hasse_vi_N4} and \ref{fig:hasse_bl_N4}, we can determine the closest partitions to any $\bc$. For example, the closest partitions to $\1$ are the partitions which split $\1$ into two clusters, one singleton and one containing all other observations; and the closest partitions to $(\lbrace 1 \rbrace, \lbrace 2 \rbrace, \lbrace 3,4 \rbrace)$ are the partition which merges the two smallest clusters $(\lbrace 1, 2 \rbrace, \lbrace 3,4 \rbrace)$ and the partition which splits the cluster of size two $(\lbrace 1 \rbrace, \lbrace 2 \rbrace, \lbrace 3 \rbrace, \lbrace 4 \rbrace)$.

\begin{property}\label{prop:closest_part}
For both metrics $\VI$ and $\tB$, the closest partitions to a partition $\bc$ are:
\begin{itemize}
\item if $\bc$ contains at least two clusters of size one and at least one cluster of size two, the partitions which merge any two clusters of size one and the partitions which split any cluster of size two.
\item if $\bc$ contains at least two clusters of size one and no clusters of size two, the partitions which merge any two clusters of size one.
\item if $\bc$ contains at most one cluster of size one, the partitions which split the smallest cluster of size greater than one into a singleton and a cluster with the remaining observations of the original cluster. 
\end{itemize}
\end{property}
A proof can be found in the Supplementary Material. This property characterizes the set of estimated partitions which are given the smallest loss. Under both loss functions, the smallest loss of zero occurs when the estimated partition is equal to the truth. Otherwise, the smallest loss occurs when the estimated clustering differs from the truth by merging two singleton clusters or splitting a cluster of size two, or, if neither is possible, splitting the smallest cluster of size $n$ into a singleton and a cluster of size $n-1$. We further note that the loss of estimating the true clustering with a clustering which merges two singletons or splits a cluster of size two, is $\frac{2}{N}$ and $\frac{2}{N^2}$ for VI and $\tB$ respectively, which converges to $0$ as $N \rightarrow \infty$ for both metrics, but at a faster rate for $\tB$.

Next, we note that the Hasse diagram stretched by $\tB$ in Figure \ref{fig:hasse_bl_N4} appears asymmetric, in the sense that $\1$ is more separated from the others when compared to the Hasse diagram stretched by $\VI$ in Figure \ref{fig:hasse_vi_N4}.

\begin{property} \label{prop:symmetry}
Suppose $N$ is divisible by $k$, and let $\bc_k$ denote a partition with $k$ clusters of equal size $N/k$.
$$ \tB(\1, \bc_k)= 1-\frac{1}{k} > \frac{1}{k}-\frac{1}{N}= \tB(\0,\bc_k).$$
$$ \VI(\1, \bc_k)= \log(k)  \leq \log(N)-\log(k) =\VI(\0, \bc_k), \quad \text{for } k \leq \sqrt{N},$$
and
$$ \VI(\1, \bc_k)= \log(k)  \geq \log(N)-\log(k) =\VI(\0, \bc_k), \quad \text{for } k \geq \sqrt{N}.$$
\end{property}

Property \ref{prop:symmetry} reflects the asymmetry apparent in Figure \ref{fig:hasse_bl_N4}. In particular, for $\tB$, a partition with two clusters of equal size $\bc_2$ will always be closer to the extreme $\0$ of each data point in its own cluster than the extreme $\1$ of everyone in one cluster. However, as the sample size increases, $\bc_2$ becomes equally distant between the two extremes. For all other values of $k$, the extreme $\0$ will always be closer. This behavior is counter-intuitive for a loss function on clusterings. 
VI is much more sensible in this regard. If $k=\sqrt{N}$, $\0$ and $\1$ are equally good estimates of $\bc_k$. For $k < \sqrt{N}$, $\bc_k$ is better estimated by $\1$ and for $k> \sqrt{N}$, $\bc_k$ is better estimated by $\0$; as the sample size increases, these preferences become stronger. In particular, note that loss of estimating $\bc_2$ with $\1$ will always be smaller than estimating it with $\0$ for $N >4$.

Additionally, we observe from Figure \ref{fig:hasse_bl_N4} that the partitions with two clusters of sizes one and three are equally distant between the two extremes under $\tB$. The following property generalizes this observation.

\begin{property}
Suppose $N$ is an even and square integer. Then, the partitions with two clusters of sizes $n=\frac{1}{2}(N-\sqrt{N})$ and $N-n$ are equally distant from $\1$ and $\0$ under $\tB$.
\end{property}

This property is unappealing for a loss function, as it states that the loss of estimating a partition consisting of two clusters of sizes $\frac{1}{2}(N-\sqrt{N})$ and $\frac{1}{2}(N+\sqrt{N})$ with the partition  of only one cluster or with the partition of all singletons is the same. Intuitively, however, $\1$ is a better estimate. The behavior of VI is much more reasonable, as partitions with two clusters will always be better estimated by $\1$ than $\0$ for $N>4$ and partitions with $\sqrt{N}$ clusters of equal size are equally distant from $\0$ and $\1$.

Finally, we note that as both VI and $\tB$ are metrics on the space of clusterings, we can construct a ball around $\bc$ of size $\epsilon$, defined as:
$$ B_\epsilon(\bc)=\lbrace \bchat \in \bC : d(\bc,\bchat) \leq \epsilon \rbrace.$$
From Property \ref{prop:closest_part}, the smallest non-trivial ball will be the same for the two metrics. When considering the next smallest ball, differences emerge; a detailed example is provide in the Supplementary Material. In the authors' opinions, the VI ball more closely reflects our intuition of the closest set of partitions to $\bc$.

\section{Point estimation via the variation of information}\label{sec:pe_VI}

As detailed in the previous section, both VI and $\tB$ share several desirable properties including being aligned with the lattice of partitions and coinciding in the smallest non-trivial ball around any clustering. However, in our comparison, differences also emerged. Particularly, we find that $\tB$ exhibits some peculiar asymmetries,  preferring to split clusters over merging, and we find that the VI ball more closely reflects our intuition of the neighborhood of a partition. In light of this, we propose to use $\VI$ as a loss function in Bayesian cluster analysis. Under the VI, the optimal partition $\bc^*$ is 
\begin{align}
&\bc^*= \argmin_{\bchat} \E[ \VI(\bc,\bchat)|\mathcal{D}] \nonumber \\
&= \argmin_{\bchat} \sum_{n=1}^N \log( \sum_{n'=1}^N  \1(\widehat{c}_{n'}=\widehat{c}_n))-2\sum_{n=1}^N \E [ \log(\sum_{n'=1}^N  \1(c_{n'}=c_n, \widehat{c}_{n'}=\widehat{c}_n))|\mathcal{D} ], \label{eq:VI_tomin}
\end{align}
with $\mathcal{D}$ denoting the data.
For a given $\bchat$, the second term in (\ref{eq:VI_tomin}) can be approximated based on the MCMC output, and evaluating this term is of order $O(MN^2)$ (recall $M$ is the number of MCMC samples). This may be computationally demanding if the number of MCMC samples is large and if (\ref{eq:VI_tomin}) must be evaluated for a large number of $\bchat$. Alternatively, one can use Jensen's inequality, swapping the log and expectation, to obtain a lower bound on the expected loss which is computationally more efficient to evaluate:
\begin{align}
\argmin_{\bchat} \sum_{n=1}^N \log( \sum_{n'=1}^N \1(\widehat{c}_{n'}=\widehat{c}_n))-2\sum_{n=1}^N  \log( \sum_{n'=1}^N  P(c_{n'}=c_n|\mathcal{D}) \1(\widehat{c}_{n'}=\widehat{c}_n)) . \label{eq:approxVI_tomin}
\end{align}
Similar to minimization of the posterior expected Binder's loss, minimization of (\ref{eq:approxVI_tomin}) only depends on the posterior through the posterior similarity matrix, which can be pre-computed based on the MCMC output. In this case, computational complexity for a given $\bchat$ is reduced to $O(N^2)$.

Due to the huge dimensions of the partition space, computing the lower bound in (\ref{eq:approxVI_tomin}) for every possible $\bchat$ is practically impossible. A simple technique to find the optimal partition $\bc^*$ restricts the search space to some smaller space of partitions. The \textbf{\textsf{R}} package 'mcclust' (\cite{mcclust}), which contains tools for point estimation in Bayesian cluster analysis and cluster comparison, includes a function \texttt{minbinder()} that finds the partition minimizing the poster expected Binder's loss among the subset of partitions 1) visited in the MCMC chain or 2) explored in a hierarchical clustering algorithm with a distance of $1- P(c_n=c_{n'}|\mathcal{D})$ and average or complete linkage. An alternative search algorithm developed in  \cite{Lau07}, which is based on binary integer programming, is also implemented.

We propose a greedy search algorithm to locate the optimal partition $\bc^*$ based on the Hasse diagram, which can be used for both VI and $\tB$. In particular, given some partition $\bchat$, we consider the $l$ closest partitions that cover $\bchat$ and the $l$ closest partitions that $\bchat$ covers. Here, the distance used to determine the closest partitions corresponds to the selected loss of VI or $\tB$. Next, the posterior expected loss $\E[L(\bc,\widehat{\bchat})|\mathcal{D}]$ is computed for all proposed partitions $\widehat{\bchat}$, and we move in the direction of minimum posterior expected loss, that is the partition $\bc'$ with minimal $\E[L(\bc,\bc')|\mathcal{D}]$ is selected. The algorithm stops when no reduction in the posterior expected loss is obtained or when a maximum number of iterations has been reached. At each iteration, the computational complexity is $O(lN^2)$. 

%

We have developed an \textbf{\textsf{R}} package 'mcclust.ext' (\cite{mcclust.ext}), expanding upon the 'mcclust' package, that is currently available on the author's website\footnote{\wade} and includes functions \texttt{minbinder.ext()} and \texttt{minVI()} to find the partition minimizing the posterior expected Binder's loss and VI, respectively. In addition to implementing the search algorithms of  \texttt{minbinder()} in 'mcclust' described previously, the greedy search algorithm is also included. As is common in greedy search algorithms, results are sensitive to both the starting value of $\bchat$ and the step size $l$. In practice, we recommend multiple restarts, for example, at different MCMC samples or the best partition found by the other search algorithms. A larger value of $l$ will allow more exploration and reduce the need for multiple restarts, and we have chosen a default value of $l=2N$ as this showed good exploration in the examples considered with little sensitivity to the initial value of $\bchat$. However, for larger datasets, this may be too expensive and multiple restarts with smaller $l$ may be preferred.

An advantage of the greedy search algorithm over simply restricting to partitions visited in the chain is that partitions not explored in the MCMC algorithm can be considered; in fact, in almost all simulated and real examples, 
the clustering estimate is not among the sampled partitions and results in a lower expected loss than any sampled partition. 

\section{Credible balls of partitions}\label{sec:CB}

To characterize the uncertainty in the point estimate $\bc^*$, we propose to construct a credible ball of a given credible level $1-\alpha$, $ \alpha \in [0,1]$, defined as 
$$ B_{\epsilon^*}(\bc^*)=\lbrace \bc : d(\bc^*,\bc)\leq \epsilon^* \rbrace,$$
where $\epsilon^*$ is the smallest $\epsilon \geq 0$ such that
$ P(B_\epsilon(\bc^*)|\mathcal{D}) \geq 1-\alpha.$
The credible ball is the smallest ball around $\bc^*$ with posterior probability at least $1-\alpha$. It reflects the posterior uncertainty in the point estimate $\bc^*$; with probability $1-\alpha$, we believe that the clustering is within a distance of $\epsilon^*$ from the point estimate $\bc^*$ given the data. It can be defined based on any metric on the space of partitions, such as VI and $\tB$. If the smallest non-trivial ball under VI or $\tB$ has posterior probability of at least $1-\alpha$, the credible balls under the two metrics will coincide (see Property \ref{prop:closest_part}). Typically, however, they will be different. 

From the MCMC output, we can obtain an estimate of $\epsilon^*$, and thus the credible ball of level $1- \alpha$. First, the distance between all MCMC samples $\lbrace \bc^m \rbrace$ and $\bc^*$ is computed. For any $\epsilon \geq 0$,
$$ P(B_\epsilon(\bc^*)|\mathcal{D})=\E[\1(d(\bc^*,\bc) \leq \epsilon)| \mathcal{D} ] \approx \frac{1}{M} \sum_{m=1}^{M} \1(d(\bc^*,\bc^m) \leq \epsilon),$$
and  $\epsilon^*$ is the smallest $\epsilon \geq 0$ such that
$ \frac{1}{M} \sum_{m=1}^{M} \1(d(\bc^*,\bc^m) \leq \epsilon) \geq 1-\alpha. $

To characterize the credible ball, we define the \textit{vertical} and \textit{horizontal bounds} of the credible ball. The vertical upper bounds consist of the partitions in the credible ball with the smallest number of clusters that are most distant from $\bc^*$. The vertical lower bounds consist of the partitions in the credible ball with the largest number of clusters that are most distant from $\bc^*$. The horizontal bounds consist of the partitions in the credible ball that are most distant from $\bc^*$. The bounds are defined more formally below, where the notation $k(\bc)$ is used for the number of clusters in $\bc$.
\begin{definition}[Vertical upper bounds] The vertical upper bounds of the credible ball $ B_{\epsilon^*}(\bc^*)$, denoted $v^u_{\epsilon^*}(\bc^*)$, are defined as
\begin{align*} v^u_{\epsilon^*}(\bc^*)= \lbrace \bc \in B_{\epsilon^*}(\bc^*) &:  k(\bc)\leq k(\bc') \:  \forall \, \bc' \in  B_{\epsilon^*}(\bc^*)  \text{ and } \\
&d(\bc,\bc^*)\geq d(\bc'',\bc^*) \:  \forall \, \bc'' \in  B_{\epsilon^*}(\bc^*) \text{ with } k(\bc)=k(\bc'') \rbrace. 
\end{align*}
\end{definition}
\begin{definition}[Vertical lower bounds] The vertical lower bounds of the credible ball $ B_{\epsilon^*}(\bc^*)$, denoted $v^l_{\epsilon^*}(\bc^*)$, are defined as
\begin{align*} v^l_{\epsilon^*}(\bc^*)= \lbrace \bc \in B_{\epsilon^*}(\bc^*) &:  k(\bc)\geq k(\bc') \:  \forall \, \bc' \in  B_{\epsilon^*}(\bc^*)  \text{ and } \\
&d(\bc,\bc^*)\geq d(\bc'',\bc^*) \:  \forall \, \bc'' \in  B_{\epsilon^*}(\bc^*) \text{ with } k(\bc)=k(\bc'') \rbrace. 
\end{align*}
\end{definition}
\begin{definition}[Horizontal bounds] The horizontal bounds of the credible ball $ B_{\epsilon^*}(\bc^*)$, denoted $h_{\epsilon^*}(\bc^*)$, are defined as
\begin{align*} h_{\epsilon^*}(\bc^*)= \lbrace \bc \in B_{\epsilon^*}(\bc^*):  d(\bc,\bc^*)\geq d(\bc',\bc^*) \:  \forall \, \bc' \in  B_{\epsilon^*}(\bc^*)  \rbrace. 
\end{align*}
\end{definition}
These bounds describe the extremes of the credible ball and with $1-\alpha$ posterior probability, how different we believe the partition may be from $\bc^*$. An example is provided in the Supplementary Material. In practice, we define the vertical and horizontal bounds based on the partitions in the credible ball with positive estimated posterior probability.

In existing literature, quantification of uncertainty in the clustering structure is typically described through a heat map of the estimated posterior similarity matrix. However, as opposed to the credible ball of Bayesian confidence level $1-\alpha$, there is no precise quantification of how much uncertainty is represented by the posterior similarity matrix. Moreover, in the examples of Section \ref{sec:examples}, we find that in a comparison with the 95\% credible balls, the uncertainty is under-represented by the posterior similarity matrix. Additionally, the credible balls have the added desirable interpretation of characterizing the uncertainty around the point estimate $\bc^*$.

\section{Examples}\label{sec:examples}

We provide both simulated and real examples to compare the point estimates from VI and Binder's loss and describe the credible ball representing uncertainty in the clustering estimate. 

\subsection{Simulated examples}\label{subsec:simulated}

\begin{figure}[!h]
\centering
\subfloat[Example 1: 4 clusters]{\includegraphics[scale=.4]{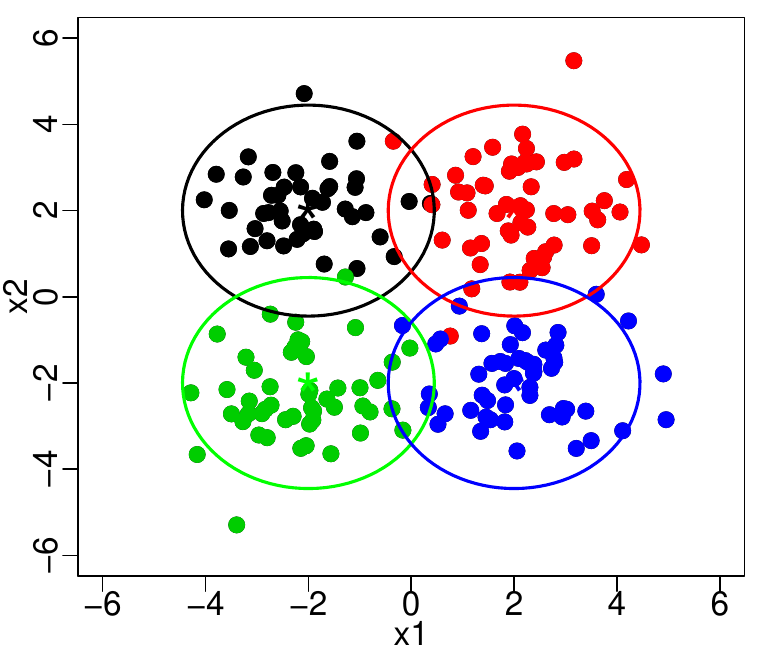} \label{fig:data_ex1}}
\subfloat[Example 2: 4 clusters]{\includegraphics[scale=.4]{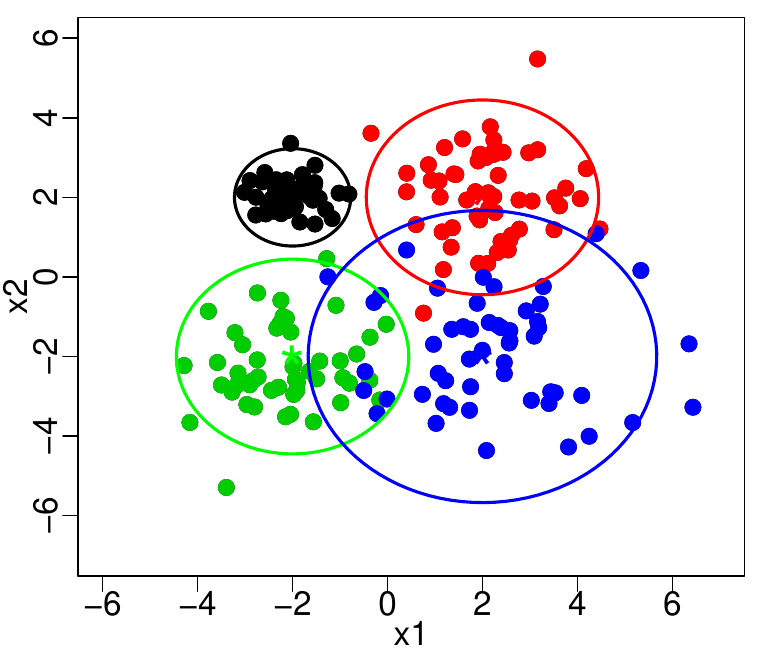}\label{fig:data_ex2}}
\caption{The data is simulated from a mixture of four normals with locations $(\pm 2, \pm 2)'$ and colored by cluster membership. In (b) components having varying standard deviations. 
}
\label{fig:data}
\end{figure}

\begin{figure}[!h]
\centering
\subfloat[Ex 1 Binder's: 9 clusters]{\includegraphics[scale=.4]{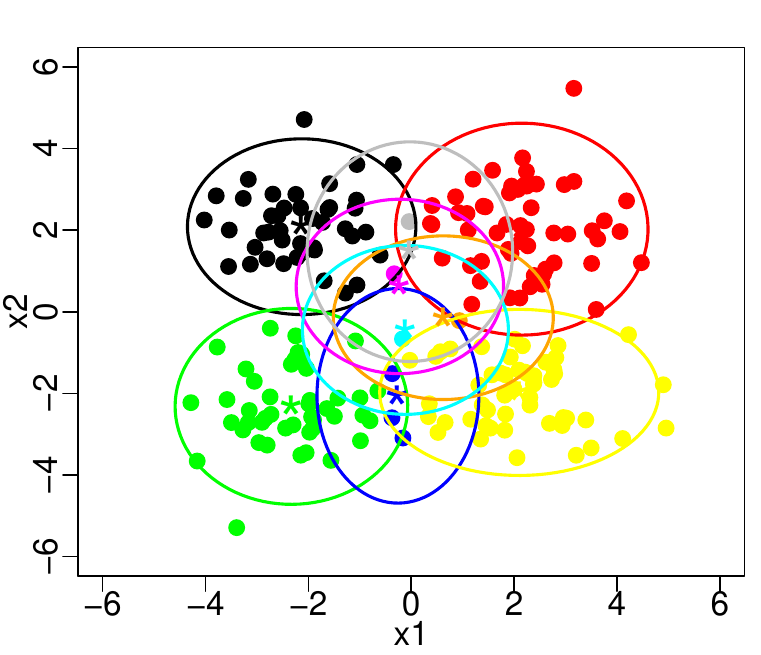}\label{fig:BL_estimate1}}
\subfloat[Ex 2 Binder's: 12 clusters]{\includegraphics[scale=.4]{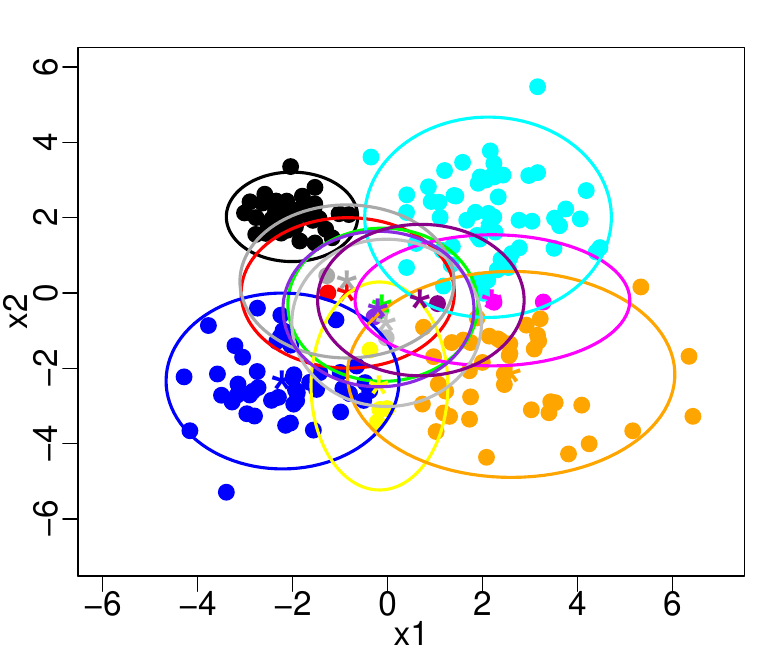}\label{fig:BL_estimate2}}\\
\subfloat[Ex 1 VI: 4 clusters]{\includegraphics[scale=.4]{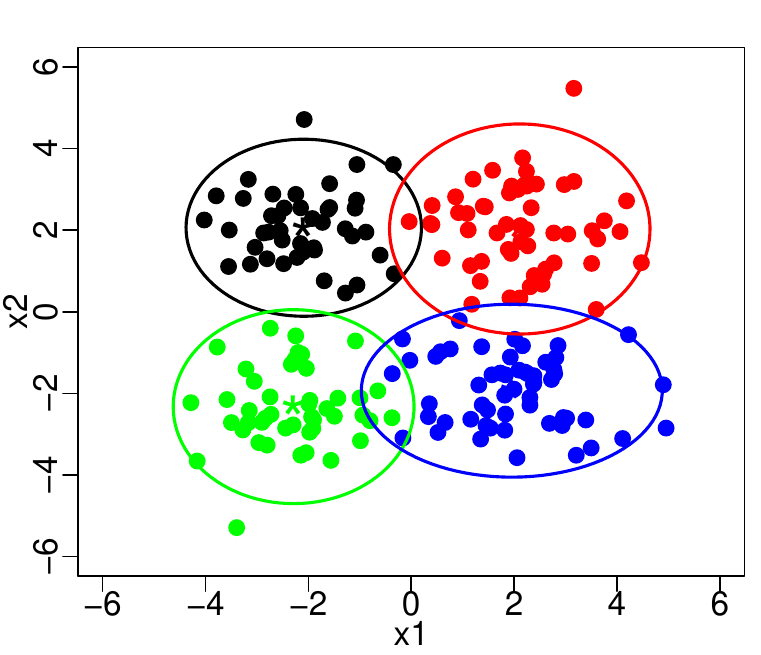}\label{fig:VI_estimate1}}
\subfloat[Ex 2 VI: 4 clusters]{\includegraphics[scale=.4]
{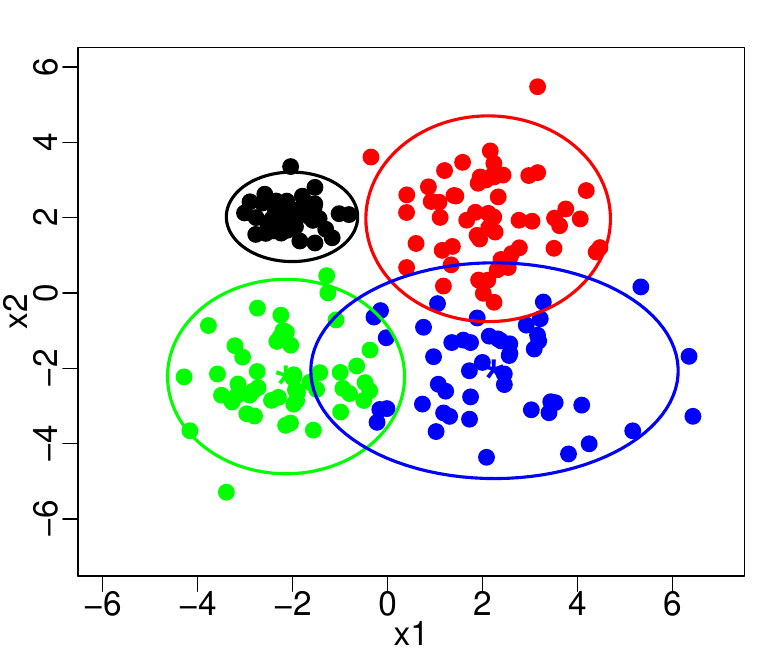}\label{fig:VI_estimate2}}
\caption{Clustering estimate with color representing cluster membership for Binder's loss (first row) and VI (second row) with columns corresponding to examples. 
}
\label{fig:ex_estimate}
\end{figure}

Two datasets of size $n=200$ are simulated from:
\begin{align*}
X_i \iidsim \sum_{j=1}^4 \frac{1}{4} \N \left( \left[\begin{array}{c} (-1)^{\lfloor \frac{(j-1)}{2} \rfloor} 2 \\ (-1)^{j-1}2 \end{array}\right], \left[\begin{array}{cc} \sigma_j^2 &0 \\ 0 & \sigma_j^2 \end{array}\right] \right). 
\end{align*}
In the first example, $\sigma_j=1$ for all components, 
while in the second example, components have varying standard deviations; $\sigma_j=1$ for the two components located in the first and third quadrants, $\sigma_j=0.5$ in the second quadrant, and $\sigma_j=1.5$ in the fourth quadrant. 
The datasets for both examples are depicted in Figure \ref{fig:data} and colored by cluster membership.


We consider a Dirichlet process (DP) mixture model:
\begin{align}
X_i|P \iidsim \int \N \left( \left[\begin{array}{c} \mu_1 \\ \mu_2 \end{array}\right], \left[\begin{array}{cc} \sigma_1^2 &0 \\ 0 & \sigma_2^2 \end{array}\right] \right) dP(\mu, \Sigma) \quad \text{and} \quad P \sim \DP(\alpha P_0)\label{eq:DPM}
\end{align}
where $ \mu= ( \mu_1, \mu_2)'$ and $\Sigma$ is a diagonal matrix with diagonal elements $(\sigma_1^2, \sigma_2^2)$. The base measure of the DP is the conjugate product of normal inverse gamma priors with parameters $(\mu_{0,i},c_{i},a_i,b_i)$ for $i=1,2$, i.e. $P_0$ has density
$$ p_0(\mu_1,\mu_2,\sigma_1^2,\sigma_2^2) \propto \prod_{i=1}^2 \sqrt{\frac{c_i}{\sigma_i^2}} \exp\left( -\frac{c_i}{2 \sigma_i^2} (\mu_i-\mu_{0,i})^2 \right) (\sigma_i^2)^{-a_i-1} \exp \left(- \frac{b_i}{\sigma_i^2} \right) .$$
The parameters were fixed to $\mu_{0,i}=0,\, c_{i}=1/2, \,a_i=2, \, b_i=1$ for $i=1,2$. The mass parameter $\alpha$ is given a $\Gam(1,1)$ hyperprior.


A marginal Gibbs sampler is used for inference (\cite{Neal}) with 10,000 iterations after a burn in period of 1,000 iterations. Trace plots and autocorrelation plots (not shown) suggest convergence. Among partitions sampled in the MCMC, only one is visited twice and all others are visited once in the first example, while no partitions are visited more than once in the second example. 

Figure \ref{fig:ex_estimate} depicts the partition estimate found by the greedy search algorithm for Binder's loss and VI and for both examples (with multiple restarts and the default value of $l=2N$); colors represent cluster membership with the posterior expected cluster-specific mean and variance represented through stars and ellipses, respectively. Tables 
in the Supplementary Material provide a comparison of the true partition with the estimates through a cross tabulation of cluster labels. In all examples, the four true clusters are visible; however, Binder's loss creates new small clusters for observations on the border between clusters where cluster membership is uncertain, overestimating the number of clusters. This effect is most extreme for the second example, where the fourth cluster (blue in Figure \ref{fig:data_ex2}) has increased overlap with the second and third clusters (red and green in Figure \ref{fig:data_ex2}), while the first cluster (black in Figure \ref{fig:data_ex2}) with decreased variance is well separated from the other clusters and identified in both estimates. 

\begin{table}[!h]
\begin{tabular}{cc|ccccccc}
  & Loss & $k_N^*$ & $N_I$ &$\E[\tB|\mathcal{D}]$ & $\tB(\bc_{t},\bc^*)$ & $\E[\VI_{\text{LB}}|\mathcal{D}]$& $\E[\VI|\mathcal{D}]$ & $\VI(\bc_{t},\bc^*)$ \\ \hline
\multirow{ 2}{*}{Ex 1:} &$\tB$ &9 &13 & \textbf{0.062} & 0.045& 0.545 & 0.816 &  0.643\\ 
&$\VI$ &\textbf{4} &\textbf{9} & 0.064 & \textbf{0.044}&\textbf{ 0.426}& \textbf{0.77} & \textbf{0.569  }\\ \hline
\multirow{ 2}{*}{Ex 2:} &$\tB$ & 12 & 18 &\textbf{0.088} & 0.056 & 0.846 & 1.068 & 0.764 \\ 
&$\VI$ & \textbf{4}&\textbf{10} & 0.093 & \textbf{0.049 }&\textbf{ 0.668}& \textbf{0.99} & \textbf{0.561} \\ \hline
\end{tabular}
\caption{A comparison of the clustering estimate with $\tB$ or VI in terms of 1) number of clusters $k_N^*$; 2) number of data points incorrectly classified, denoted $N_I$; 3) expected $\tB$; 4) $\tB$ between the optimal and true clusterings; 5) expected lower bound of VI; 6) expected VI; and 7) VI between the optimal and true clusterings for both examples.}
\label{tbl:cstar_compare}
\end{table}

\begin{table}[!h]
\begin{tabular}{cc|ccccccc}
Ex 1  & Loss & $k_N^*$& $N_I$ & $\E[\tB|\mathcal{D}]$ & $\tB(\bc_{t},\bc^*)$ & $\E[\VI_{\text{LB}}|\mathcal{D}]$& $\E[\VI|\mathcal{D}]$ & $\VI(\bc_{t},\bc^*)$ \\ \hline
\multirow{ 2}{*}{$N=200$:} &$\tB$ & 9& 13 &\textbf{0.062} & 0.045& 0.545 & 0.816 &  0.643\\ 
&$\VI$ & \textbf{4}& \textbf{9}  &0.064 & \textbf{0.044}&\textbf{ 0.426}& \textbf{0.77} & \textbf{0.569  } \\ \hline
\multirow{ 2}{*}{$N=400$:} &$\tB$ & 17& 31 &\textbf{0.068} & 0.052 & 0.674 & 1.0 & 0.769 \\ 
&$\VI$ &\textbf{4} &\textbf{18}& 0.073 & \textbf{0.044 }&\textbf{ 0.505}& \textbf{0.933} & \textbf{0.54} \\ \hline
\multirow{ 2}{*}{$N=800$:} &$\tB$ & 24& 62&\textbf{0.068} &  0.061& 0.615 & 1.016 &  0.903\\ 
& $\VI$ & \textbf{4}&\textbf{47} &0.069 &\textbf{ 0.056} &\textbf{0.477} &   \textbf{0.943}& \textbf{0.742} \\ \hline
\multirow{ 2}{*}{$N=1600$:} &$\tB$ &41 &93 & \textbf{0.058} & \textbf{0.044} & 0.551 & 0.898 & 0.719 \\ 
& $\VI$ & \textbf{4}& \textbf{49} &0.0596 & 0.045&\textbf{0.403}& \textbf{0.814} &  \textbf{0.629}
\end{tabular}
\caption{Example 1 with increasing sample size: a comparison of the clustering estimate with $\tB$ or VI in terms of 1) number of clusters $k_N^*$; 2) number of data points incorrectly classified, denoted $N_I$; 3) expected $\tB$; 4) $\tB$ between the optimal and true clusterings; 5) expected lower bound of VI; 6) expected VI; and 7) VI between the optimal and true clusterings.}
\label{tbl:cstar_comparen}
\end{table}

A further comparison of the true partition with the estimates under Binder's loss and VI, for both examples, is provided in Table \ref{tbl:cstar_compare}. As expected, the $\tB$ estimate and VI estimate achieve the lowest posterior expected loss for $\tB$ and VI, respectively, but interestingly, the VI estimate has the smallest distance from the truth for both $\tB$ and VI in both examples, with the greatest improvement in the second example. Furthermore, the number of incorrectly classified data points is greater for the $\tB$ estimate than the VI estimate.

Additional simulated experiments were performed to analyze the effect of increasing the sample size in the first example. The results are succinctly summarized in Table \ref{tbl:cstar_comparen}. As the sample size increases, more points are located on the border where cluster membership is uncertain. This results in an increasing number of clusters in the $\tB$ estimate (up to 41 clusters for $N=1600$), while the VI estimate contains only four clusters for all sample sizes. In both estimates, the number of incorrectly classified data points increases with the sample size, however this number is smaller for the VI estimate in all sample sizes, with the difference between this number for Binder's and VI growing with the sample size. Furthermore, the VI estimate has improved VI distance with truth and improved or comparable $\tB$ distance with truth when compared with the $\tB$ estimate.

\begin{figure}[!h]
\centering
\subfloat[$\tB$ estimate: 9 clusters]{\includegraphics[scale=.33]{EBL_clust_ex2_wellipse}\label{fig:BL_estimate_rep}}
\subfloat[$\tB$ horizontal bound:,11 clusters][$\tB$ horizontal bound:\\11 clusters]{\includegraphics[scale=.33]{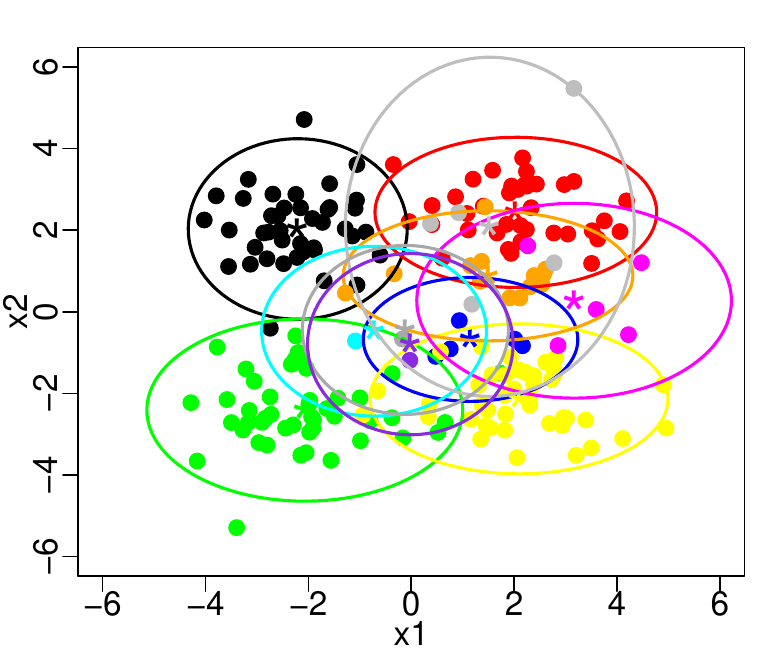}\label{fig:EBL_horiz_ex1}}\\
\subfloat[ $\tB$ upper vertical bound:, 4 clusters][$\tB$ upper vertical bound:\\4 clusters]{\includegraphics[scale=.33]{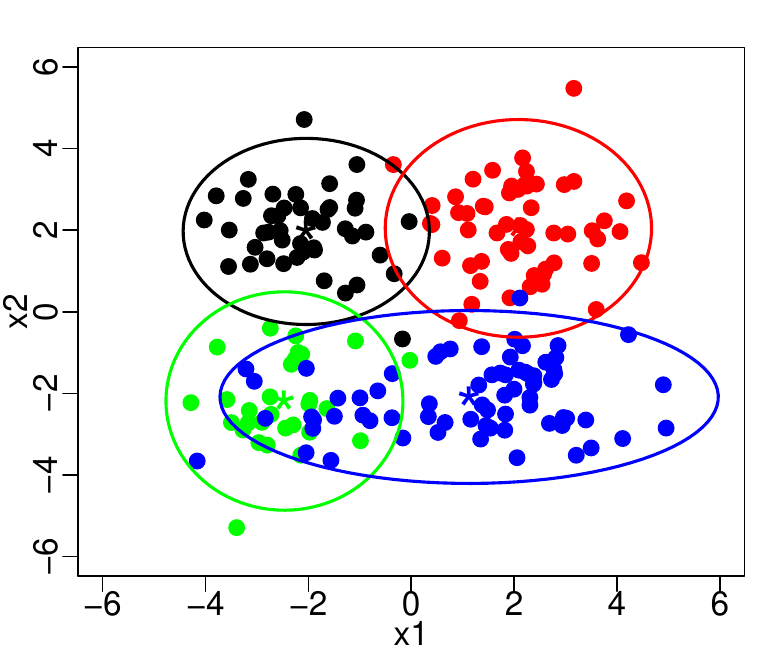}\label{fig:EBL_upper_ex1}}  
\subfloat[$\tB$ lower vertical bound:, 18 clusters][$\tB$ lower vertical bound:\\18 clusters]{\includegraphics[scale=.33]{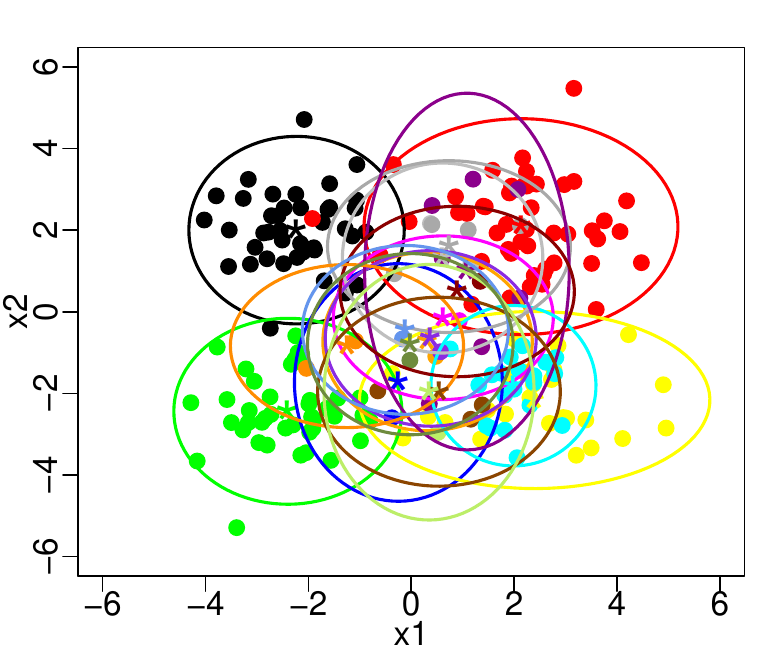}\label{fig:EBL_lower_ex1}}
\subfloat[Posterior similartiy matrix]{\includegraphics[scale=.33]{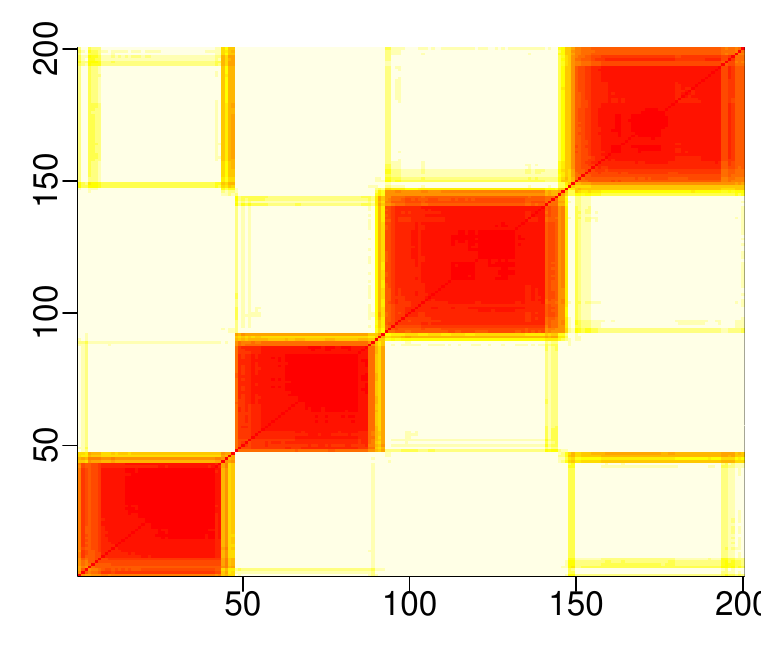}\label{fig:pmat_ex1}}
\caption{Example 1: 95\% credible ball with Binder's loss around $\bc^*$ (a) represented by the (b) horizontal bound, (c) upper vertical bound, and (d) lower vertical bound, where color denotes cluster membership, and a heat map of the posterior similarity matrix (e).}
\label{fig:EBL_bounds_ex1}
\end{figure}

\begin{table}[!h]
\begin{tabular}{cc|cc|cc|cc}
  & Loss & \multicolumn{2}{c|}{Upper} &  \multicolumn{2}{c|}{Lower} &  \multicolumn{2}{c}{Horizontal} \\
  & & $k^u_N$ & $d(\bc^*,\bc_u)$ & $k^l_N$ & $d(\bc^*,\bc_l)$ & $k^h_N$ & $d(\bc^*,\bc_h)$ \\  \hline
\multirow{ 2}{*}{Ex 1:} &$\tB$ &4 & 0.097 & 18 & 0.097 &11& 0.097\\ 
&$\VI$ & 4 & 1.02 & 16 & 1.152 & 11 & 1.213\\ \hline
\multirow{ 2}{*}{Ex 2:} &$\tB$ & 4 & 0.137 & 19 & 0.131 & 10 &0.137 \\ 
&$\VI$ & 3 & 1.043 & 16 & 1.342 & 6 & 1.403 \\ 
\end{tabular}
\caption{A summary of the credible bounds with $\tB$ or VI in terms of the number of clusters and distance to the clustering estimate for the upper vertical, lower vertical, and horizontal bounds and for both examples.}
\label{tbl:cb_compare}
\end{table}

Further experiments were carried out to consider highly unbalanced clusters. In this case, the conclusions continue to hold; Binder's loss overestimates the number of clusters present, placing uncertain observations in new small clusters, and this effect becomes more pronounced with increased overlap between clusters (results not shown).

\begin{figure}[!h]
\centering
\subfloat[VI estimate: 4 clusters]{\includegraphics[scale=.33]{VI_clust_ex2_wellipse}\label{fig:VI_estimate_rep}}
\subfloat[VI horizontal bound:, 11 clusters][VI horizontal bound:\\11 clusters]{\includegraphics[scale=.33]{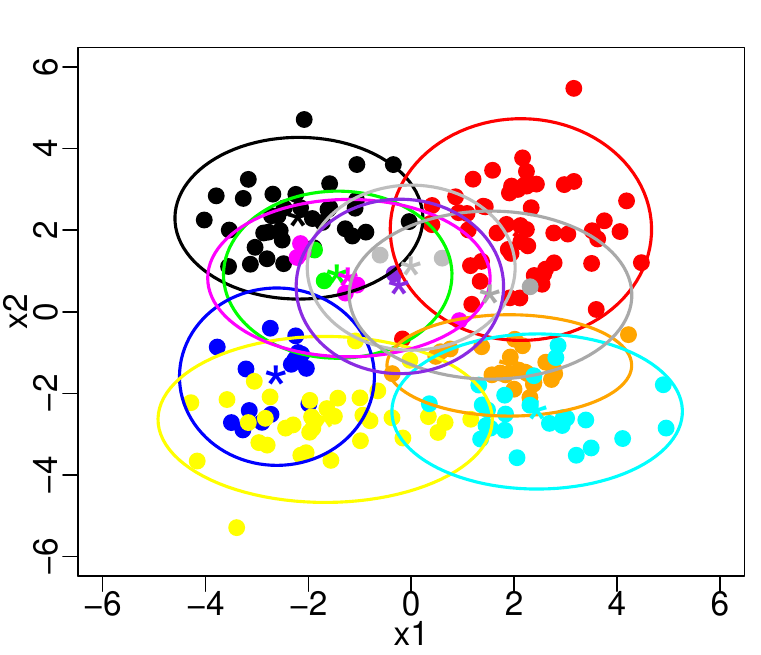}\label{fig:VI_horiz_ex1}}
\\
\subfloat[VI upper vertical bound:, 4 clusters][VI upper vertical bound:\\4 clusters]{\includegraphics[scale=.33]{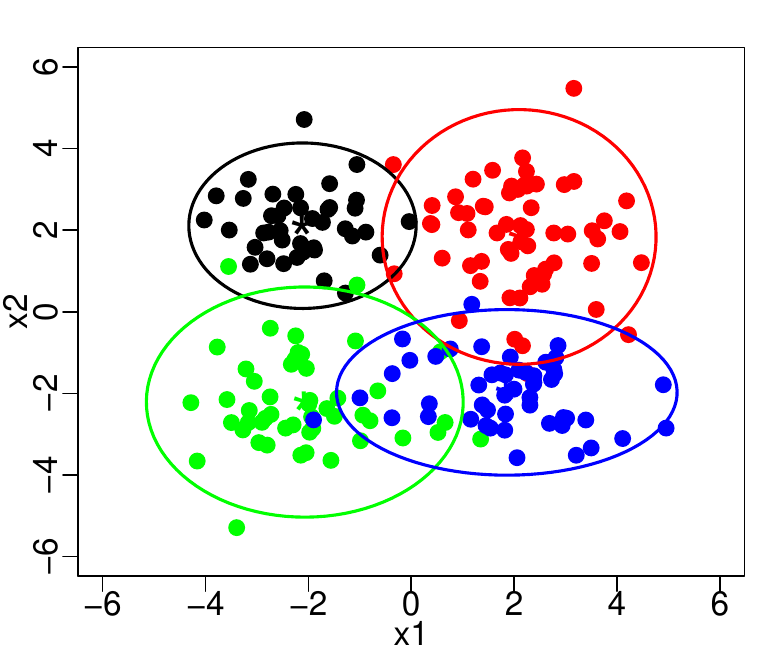}\label{fig:VI_upper_ex1}}
\subfloat[VI lower vertical bound:, 16 clusters][VI lower vertical bound:\\16 clusters]{\includegraphics[scale=.33]{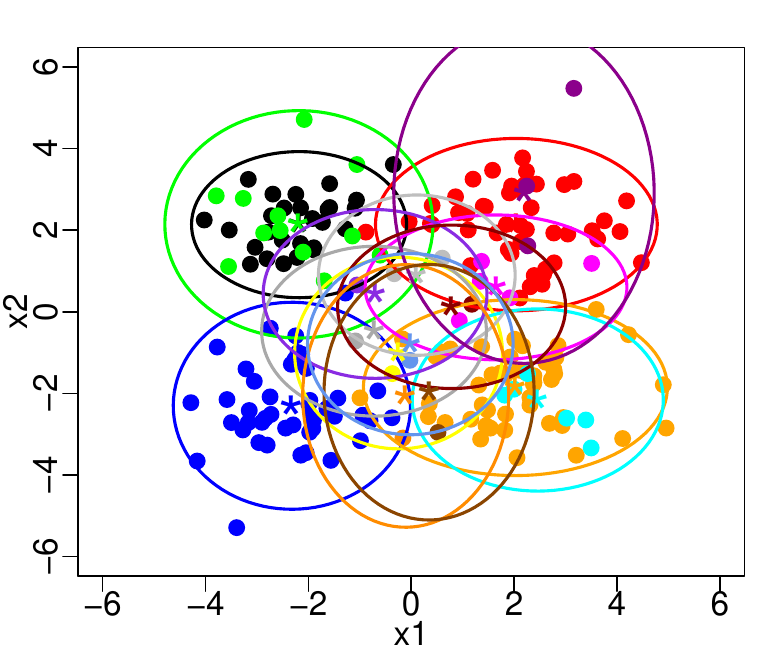}\label{fig:VI_lower_ex1}}
\subfloat[Posterior similartiy matrix]{\includegraphics[scale=.33]{Pmat_ex2_nolabels}\label{fig:pmat_ex1}}
\caption{Example 1: 95\% credible ball with VI around $\bc^*$ (a) represented by the (b) horizontal bound, (c) upper vertical bound, and (d) lower vertical bound, where color denotes cluster membership, and a heat map of the posterior similarity matrix (e).}
\label{fig:VI_bounds_ex1}
\end{figure}

\begin{figure}[!h]
\centering
\subfloat[$\tB$ estimate: 12 clusters]{\includegraphics[scale=.33]{EBL_clust_ex3_wellipse}\label{fig:BL_estimate_ex2}}
\subfloat[$\tB$ horizontal bound:, 10 clusters][$\tB$ horizontal bound:\\10 clusters]{\includegraphics[scale=.33]{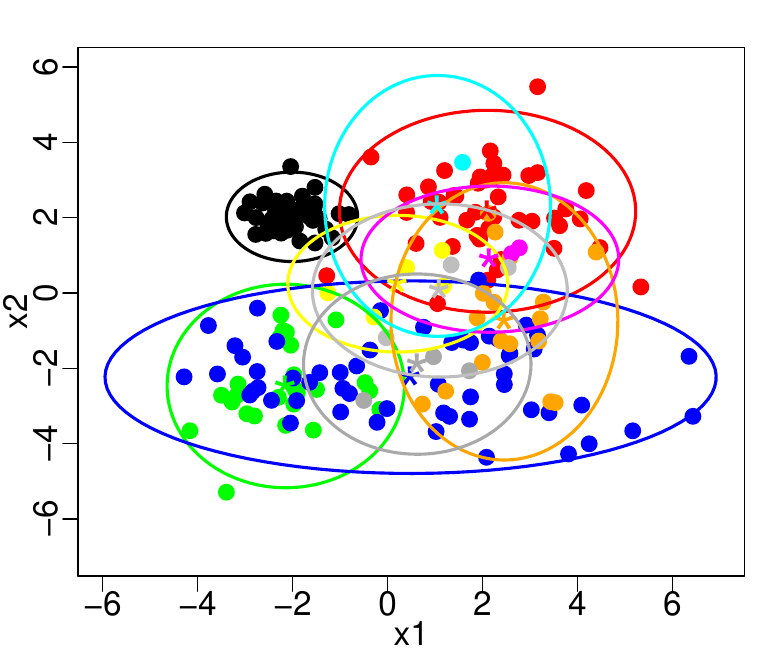}\label{fig:EBL_horiz_ex2}}\\
\subfloat[$\tB$ upper vertical bound:,4 clusters][$\tB$ upper vertical bound:\\4 clusters]{\includegraphics[scale=.33]{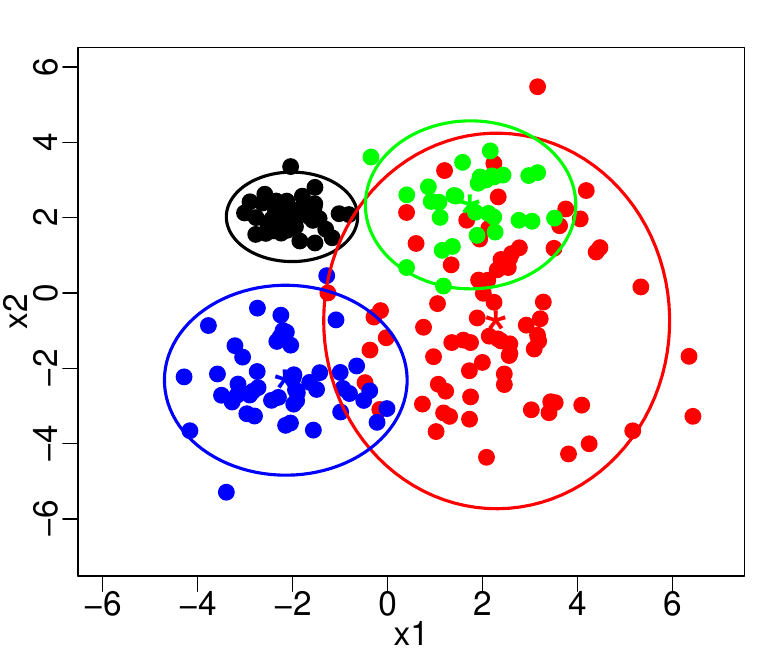}\label{fig:EBL_upper_ex2}}
\subfloat[$\tB$ lower vertical bound:,19 clusters][$\tB$ lower vertical bound:\\19 clusters]{\includegraphics[scale=.33]{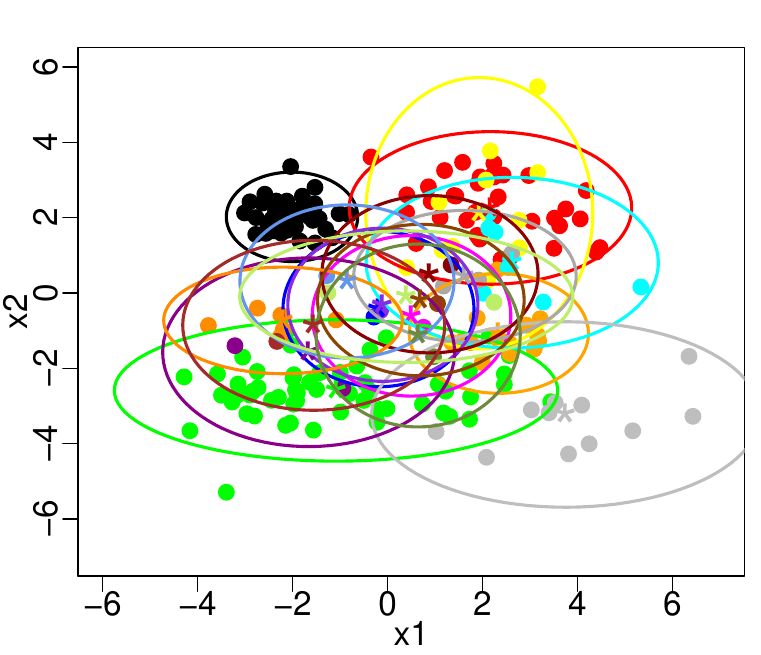}\label{fig:EBL_lower_ex2}}
\subfloat[Posterior similartiy matrix]{\includegraphics[scale=.33]{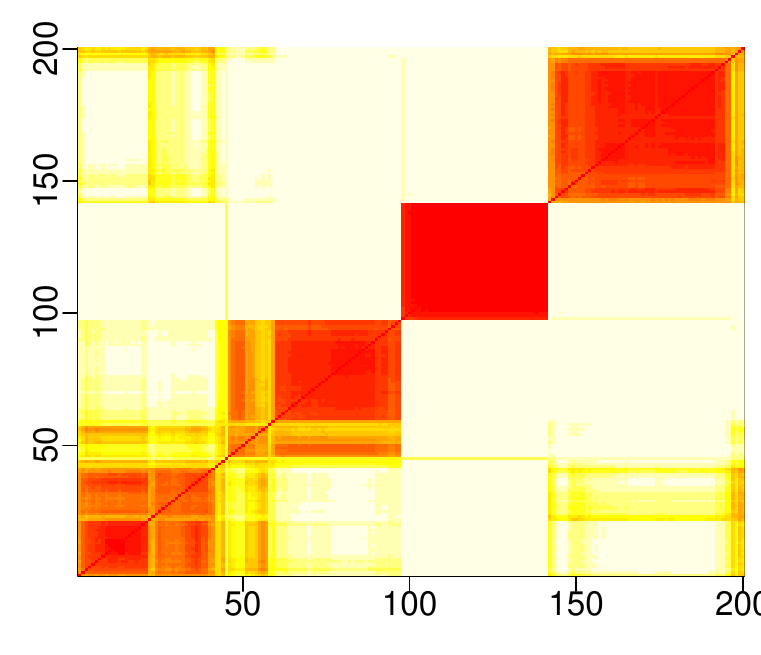}\label{fig:pmat_ex2}}
\caption{Example 2: 95\% credible ball with Binder's loss around $\bc^*$ (a) represented by the (b) horizontal bound, (c) upper vertical bounds (only one of two shown for conciseness), and (d) lower vertical bound, where color denotes cluster membership, and a heat map of the posterior similarity matrix (e).}
\label{fig:EBL_bounds_ex2}
\end{figure}

\begin{figure}[!h]
\centering
\subfloat[VI estimate: 4 clusters]{\includegraphics[scale=.33]{VI_clust_ex3_wellipse}\label{fig:VI_estimate_ex2}}
\subfloat[VI horizontal bound:, 6 clusters][VI horizontal bound:\\ 6 clusters]{\includegraphics[scale=.33]{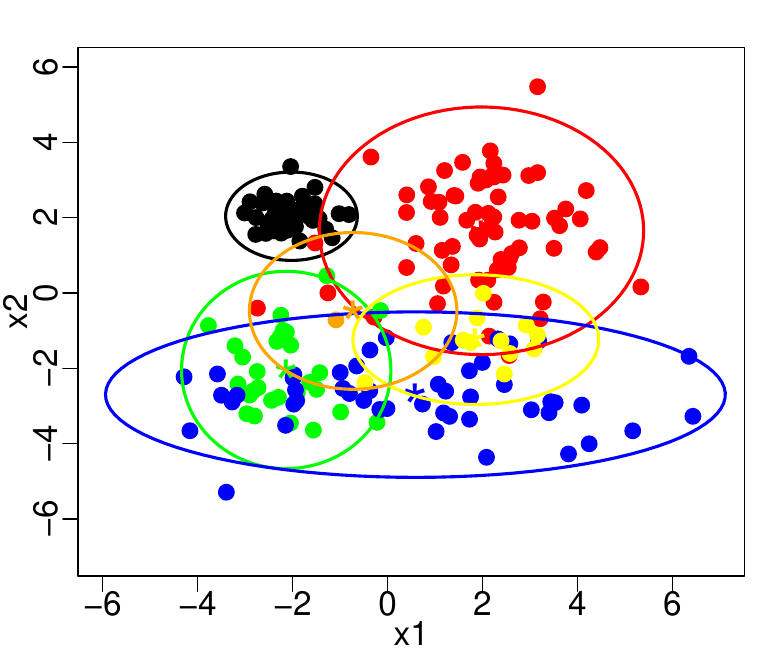}\label{fig:VI_horiz_ex2}}\\
\subfloat[VI upper vertical bound:, 3 clusters][VI upper vertical bound:\\3 clusters]{\includegraphics[scale=.33]{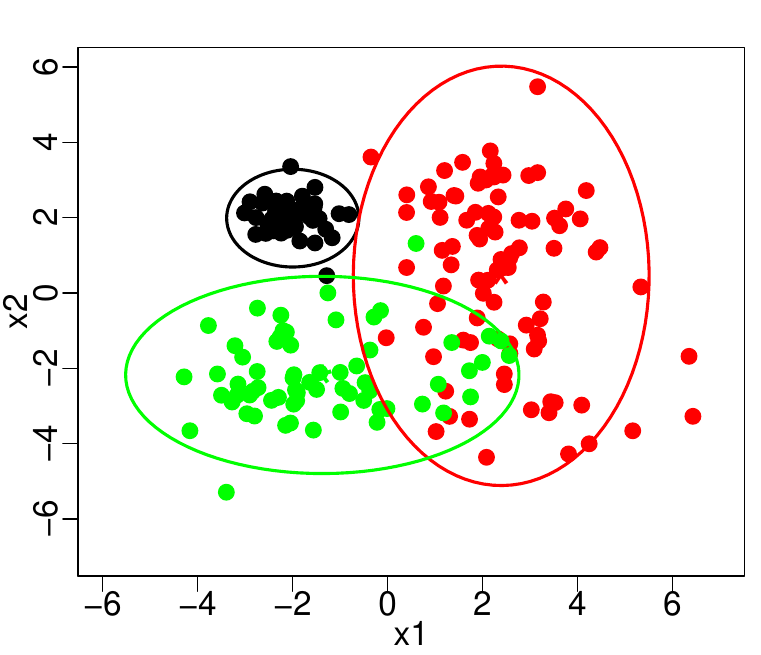}\label{fig:VI_upper_ex2}}
\subfloat[VI lower vertical bound:, 16 clusters][VI lower vertical bound:\\16 clusters]{\includegraphics[scale=.33]{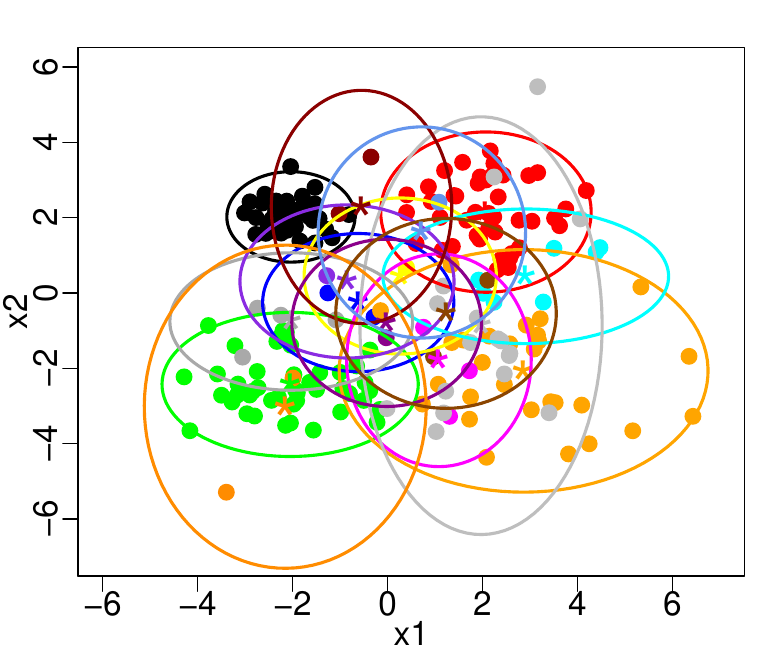}\label{fig:VI_lower_ex2}}
\subfloat[Posterior similarity matrix]{\includegraphics[scale=.33]{Pmat_ex3_nolabels}\label{fig:pmat_ex2}}
\caption{Example 2: 95\% credible ball with VI around $\bc^*$ (a) represented by the (b) horizontal bound, (c) upper vertical bound, and (d) lower vertical bound, where color denotes cluster membership and a heat map of the posterior similarity matrix (e).}
\label{fig:VI_bounds_ex2}
\end{figure}

For the first example, Figures \ref{fig:EBL_bounds_ex1} and \ref{fig:VI_bounds_ex1} represent the 95\% credible ball around the optimal partition for $\tB$ and VI, respectively, through the upper vertical bound, lower vertical bound, and horizontal bound, with data points colored according to cluster membership. Analogous plots for the second example are found in Figures \ref{fig:EBL_bounds_ex2} and \ref{fig:VI_bounds_ex2}. The Supplementary Material provides tables comparing the bounds with the true clustering through a cross tabulation of the true cluster labels with the cluster labels for each bound.

In the first example, we observe that elements of the 95\% credible ball with positive estimated posterior probability have at least four clusters for both metrics and at most 18 clusters for $\tB$ or 16 clusters for $\VI$, while the most distant elements contain 11 clusters for $\tB$ and $\VI$ (Table \ref{tbl:cb_compare}). For both metrics, these bounds reallocate uncertain data points on the border with these points either added to one of the four main clusters or to new small to medium -sized clusters. For example, in the $\tB$ upper bound, 19 elements of the third cluster (green in Figure \ref{fig:data_ex1}) are added to the fourth cluster (blue in Figure \ref{fig:data_ex1}) and in the $\tB$ lower bound, the fourth cluster (blue in Figure \ref{fig:data_ex1}) is split in two medium-sized clusters and several small clusters.


In the second example, the first cluster (black in Figure \ref{fig:data_ex2}) is stable in all bounds, while the 95\% credible ball reflects posterior uncertainty on whether to divide the remaining data points into 3 to 18 clusters for $\tB$ and 2 to 15 clusters for VI  (Table \ref{tbl:cb_compare}). Notice the high uncertainty in the fourth cluster with increased variance (blue in Figure \ref{fig:data_ex2}). Additionally, note the greater uncertainty around the optimal estimate in Example 2, as the horizontal distance in Table \ref{tbl:cb_compare} is greater for Example 2  for both metrics. 
Figures \ref{fig:EBL_bounds_ex1}-\ref{fig:VI_bounds_ex2} 
also present heat maps of the posterior similarity matrix for both examples. In general, the posterior similarity matrix appears to under-represent the uncertainty; indeed, one would conclude from the similarity matrix that there is only uncertainty in allocation of a few data points in Example 1. Moreover, the 95\% credible ball gives a precise quantification of the uncertainty. 

\subsection{Galaxy example}\label{subsec:real}

We consider an analysis of the galaxy data (\cite{Roeder90}), available in the MASS package of \textbf{\textsf{R}}, which contains measurements of velocities in km/sec of 82 galaxies from a survey of the Corona Borealis region. The presence of clusters provides evidence for voids and superclusters in the far universe. The data is modeled with a DP mixture (\ref{eq:DPM}). The parameters were selected empirically with $\mu_{0}=\bar{x},c=1/2,a=2,b=s^2$, where $\bar{x}$ represents the sample mean and $s^2$ represents the sample variance. The mass parameter $\alpha$ is given a $\Gam(1,1)$ hyperprior.

\begin{figure}[!h]
\centering
\subfloat[Binder's loss: 7 clusters]{\includegraphics[scale=.4]{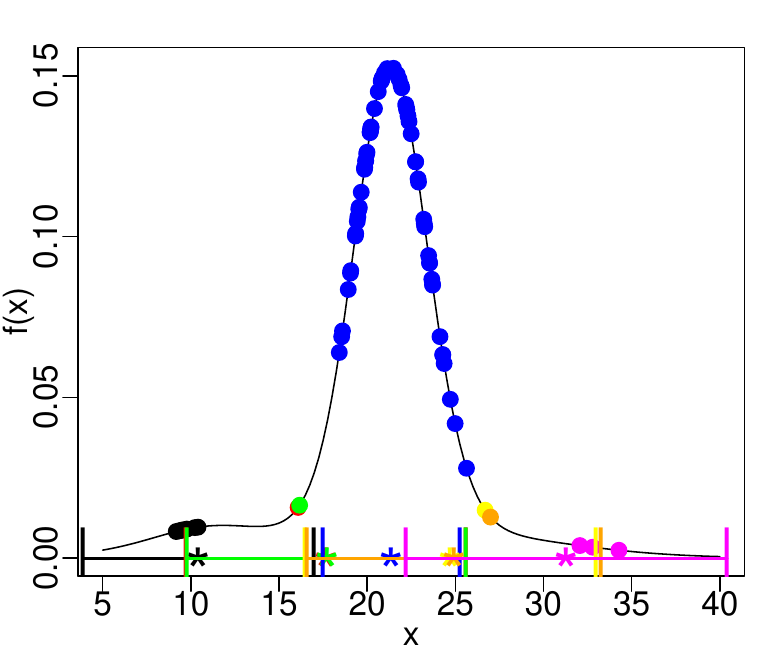}\label{fig:BL_estimate_galaxy}}
\subfloat[VI: 3 clusters]{\includegraphics[scale=.4]{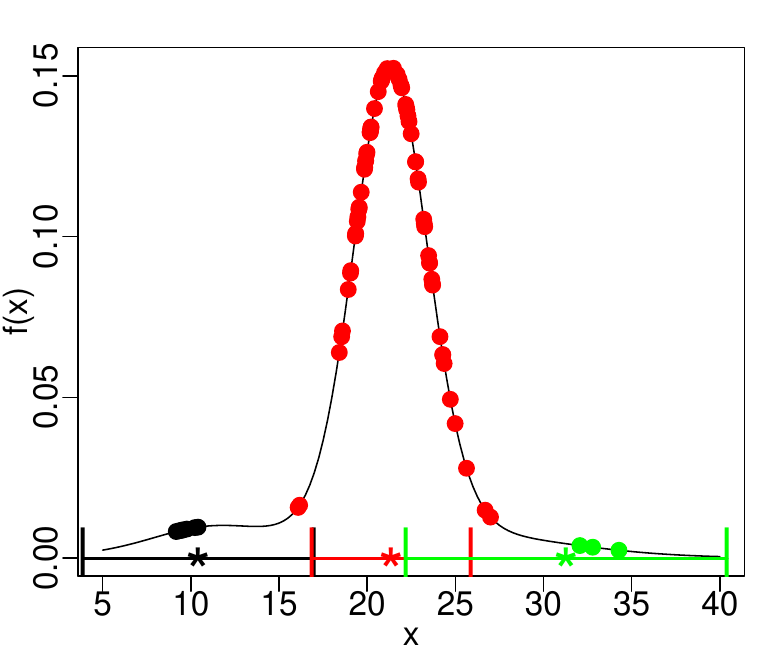}\label{fig:VI_estimate_galaxy}}
\caption{Galaxy example: optimal clustering estimate with color representing cluster membership for Binder's loss and VI, with correspondingly colored stars and bars along the x-axis representing the posterior mean and variance within cluster.  
}
\label{fig:galaxy_estimate}
\end{figure}

With 10,000 samples after 1,000 burn in, the posterior mass is spread out over 9,636 partitions, emphasizing the need for appropriate summary tools. Figure \ref{fig:galaxy_estimate} plots the point estimate of the partition found by the greedy search algorithm for Binder's loss and VI (with multiple restarts and the default value of $l=2N$). The data values are plotted against the estimated density values from the DP mixture model and colored according to cluster membership, with correspondingly colored stars and bars along the x-axis representing the posterior mean and variance within cluster. Again, we observe that Binder's loss places observations with uncertain allocation into singleton clusters, with a total of 7 clusters, 4 of which are singletons, while the VI solution contains 3 clusters. Table \ref{tbl:cstar_compare_galaxy} compares the point estimates in terms of the posterior expected $\tB$, lower bound of $\VI$, and $\VI$; as anticipated, the $\tB$ solution has the smallest posterior expected $\tB$ and the VI solution has the smallest posterior expected VI. 

\begin{table}[!h]
\begin{tabular}{c|cccc}
 Loss & $k_N^*$ & $\E[\tB|\mathcal{D}]$ & $\E[\VI_{\text{LB}}|\mathcal{D}]$ & $\E[\VI|\mathcal{D}]$ \\ \hline
 $\tB$ & 7 & \textbf{0.218} & 0.746 &1.014 \\ 
$\VI$ & 3 & 0.237 & \textbf{0.573} &\textbf{0.939} \\
\end{tabular}
\caption{Galaxy example: a comparison of the optimal partition with Binder's loss and VI  in terms of posterior expected $\tB$, lower bound to VI, and VI.}
\label{tbl:cstar_compare_galaxy}
\end{table}

\begin{figure}[!h]
\centering
\subfloat[upper vertical bound:, 2 clusters][Upper vertical bound: 2 clusters]{\includegraphics[scale=.4]{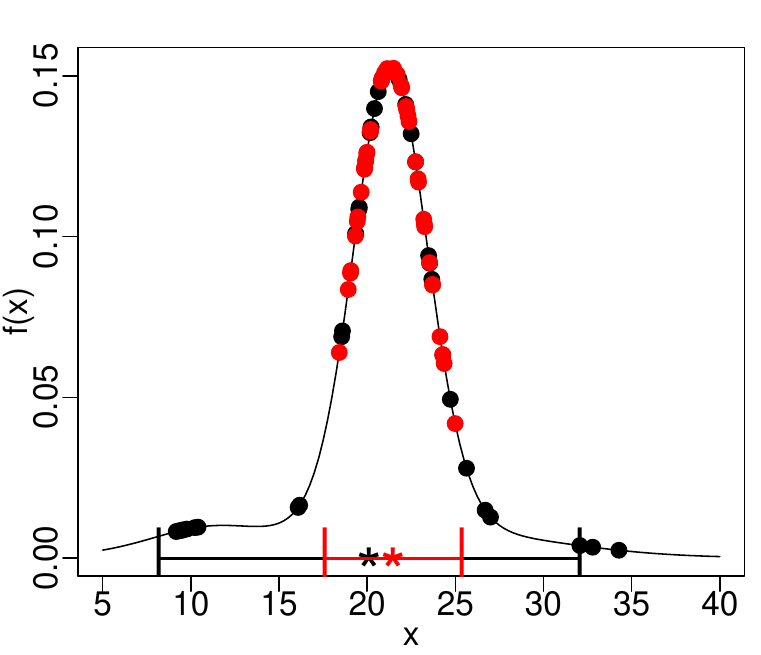}\label{fig:VI_upper_galaxy}}
\subfloat[lower vertical bound:, 15 clusters][Lower vertical bound: 15 clusters]{\includegraphics[scale=.4]{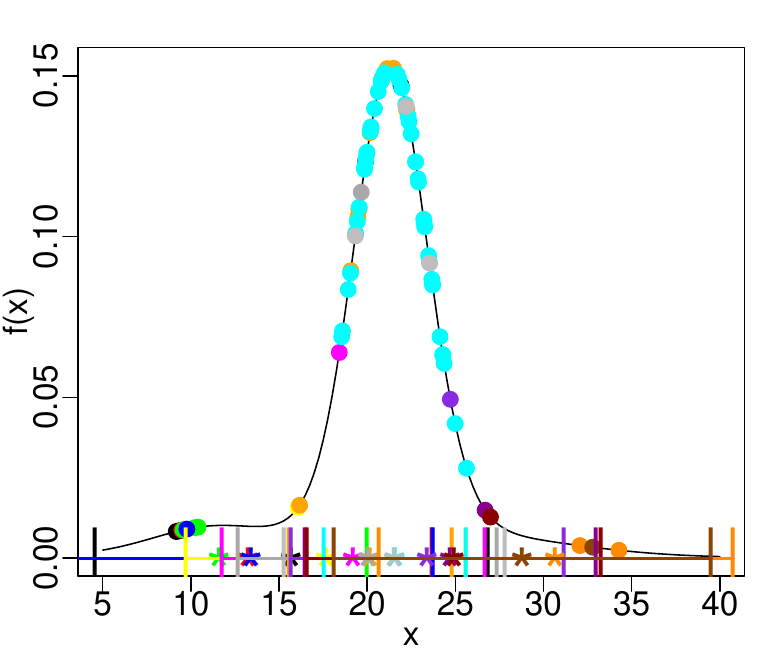}\label{fig:VI_lower_galaxy}}\\
\subfloat[$\VI$ horizontal bound, 8 clusters][Horizontal bounds: 8 clusters]{\includegraphics[scale=.4]{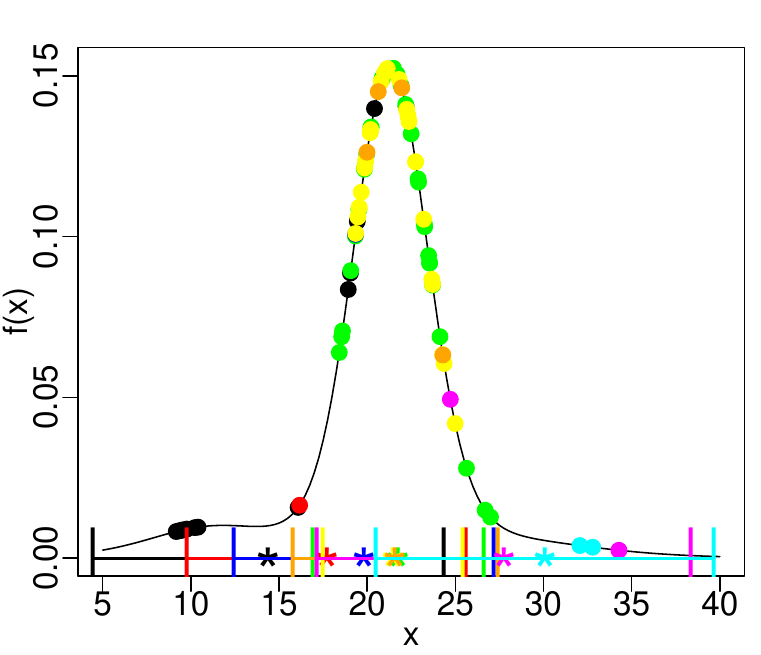}\label{fig:VI_horiz_galaxy}}
\subfloat[Posterior similarity matrix]{\includegraphics[scale=.4]{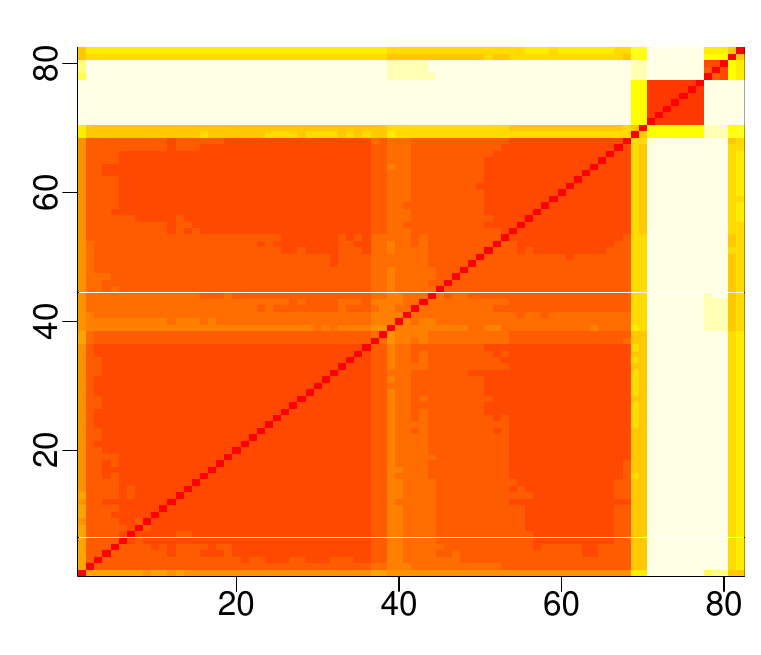}\label{fig:pmat_galaxy}}
\caption{Galaxy example: 95\% credible ball with VI represented by the (a) upper vertical bound, (b) lower vertical bound, and (c) horizontal bound, where color denotes cluster membership, with correspondingly colored stars and bars along the x-axis representing the posterior mean and variance within cluster, and (d) a heat map of the posterior similarity matrix.}
\label{fig:VI_bounds_galaxy}
\end{figure}

\begin{table}[!h]
\begin{tabular}{c|cc|cc|cc}
   & \multicolumn{2}{c|}{Upper} &  \multicolumn{2}{c|}{Lower} &  \multicolumn{2}{c}{Horizontal} \\
  & $k^u_N$ & $d(\bc^*,\bc_u)$ & $k^l_N$ & $d(\bc^*,\bc_l)$ & $k^h_N$ & $d(\bc^*,\bc_h)$ \\  \hline
Galaxy &2 & 1.364 & 15 & 1.669 &8& 1.832\\ 
\end{tabular}
\caption{Galaxy example: a summary of the credible bounds with VI in terms of the number of clusters and distance to the clustering estimate for the upper vertical, lower vertical, and horizontal bounds.}
\label{tbl:cb_compare_galaxy}
\end{table}

The 95\% 
VI credible ball contains all partitions with a VI distance less than 1.832. Figure 
\ref{fig:VI_bounds_galaxy} depicts the 95\% credible ball through the upper vertical, lower vertical, and horizontal bounds, which are further described and summarized in Table \ref{tbl:cb_compare_galaxy} and in cross tabulation tables in the Supplementary Material. 
We observe a large amount of variability around the optimal partition. With 95\% posterior probability, we believe that, on one extreme, the data could be modeled using only 2 components, one with a large variance to account for outliers (black cluster in Figure (\ref{fig:VI_upper_galaxy})). On the other extreme, the data could be further split into one medium sized cluster and many, 14 to be precise, smaller clusters. The horizontal bound, the most extreme partition in the credible ball, splits the largest cluster (red in Figure \ref{fig:VI_estimate_galaxy}) into two medium sized clusters and four small clusters and reallocates some of its data points to the first cluster (black in Figure \ref{fig:VI_estimate_galaxy}). Figure \ref{fig:pmat_galaxy} emphasizes that the posterior similarity matrix under-represents the uncertainty around the point estimate in comparison to the credible ball. 

\section{Discusssion}\label{sec:discussion}

Bayesian cluster analysis provides an advantage over classical cluster analysis, in that the Bayesian procedure returns a posterior distribution over the entire partition space, reflecting uncertainty in the clustering structure given the data, as opposed to returning a single solution or conditioning on the parameter estimates and number of clusters. This allows one to assess statistical properties of the clustering given the data. However, due to the huge dimension of the partition space, an important problem in Bayesian cluster analysis is how to appropriately summarize the posterior. To address this problem, we have developed tools to obtain a point estimate of clustering based on the posterior and describe uncertainty around this estimate via the 95\% credible ball.

Obtaining a point estimate through a formal decision theory framework requires the specification of a loss function. Previous literature focused on Binder's loss. In this work, we propose to use an information theoretic measure, the variation of information, and provide a detailed comparison of the two metrics. 
We find that Binder's loss exhibits peculiar asymmetries, preferring to split over merge clusters, and 
the variation of information is more symmetric in this regard. This behavior of Binder's loss causes the optimal partition to overestimate the number of clusters, allocating uncertain data points to small additional clusters. In addition, we have developed a novel greedy search algorithm to locate the optimal partition, allowing one to explore beyond the space of partitions visited in the MCMC chain. 

To represent uncertainty around the point estimate, we construct 95\% credible balls around the point estimate and depict the credible ball through the upper vertical, lower vertical, and horizontal bounds. In addition to a heat map of the posterior similarity matrix, which is often reported in literature, the 95\%  credible ball enriches our understanding of the uncertainty present. Indeed, it provides a precise quantification of the uncertainty present around the point estimate, and in examples, we find that an analysis based on the posterior similarity matrix leads one to be over certain in the clustering structure.The developed posterior summary tools for Bayesian cluster analysis are available\footnote{through the author's website \wade} through an \textbf{\textsf{R}} package 'mcclust.ext' (\cite{mcclust.ext}), expanding upon the existing \textbf{\textsf{R}} package 'mcclust' (\cite{mcclust}).

In future work, we aim to extend these ideas to Bayesian feature allocation analysis, an extension of clustering which allows observations to belong to multiple clusters (\cite{GG11}). A further direction of research will be to explore posterior consistency for the number of clusters based on the VI estimate for BNP mixture models; this is of particular interest in light of the negative results of \cite{MH13} 
and \cite{MH14} 
and the positive results in our simulation studies 
(Table \ref{tbl:cstar_comparen}). 
Finally, scalability issues of BNP mixture models are an important concern for very large datasets. To scale with large sample sizes, a number of papers have avoided exploration of the posterior on partitions through MCMC and focused on finding a point estimate of the partition, often through MAP inference (\cite{Heller05}, \cite{Dahl09}, \cite{RBL14}) or the DP-means algorithm and its extensions (\cite{KJ12}, \cite{JKJ12}, \cite{BKJ13}). One direction of future research is to develop an algorithm to find the point estimate which minimizes the posterior expected VI that avoids MCMC. 
Of course, while gaining in scalability, we lose the uncertainty in the clustering structure. 

\medskip

\noindent \textbf{Acknowledgements} This work was supported by the Engineering and Physical Sciences Research Council [grant number EP/I036575/1].

\bibliographystyle{plainnat}
\bibliography{CA}

\end{document}


\title{Supplementary material for Bayesian cluster analysis: Point estimation and credible balls}
\author{Sara Wade and Zoubin Ghahramani}

\maketitle

\section{Review of Lattice Theory}

The space of partitions is a partially ordered set. In particular, equip the space of partitions $\bC$ with the binary relation $\leq$ on $\bC$ defined by set containment, i.e. for $ \bc, \bchat \in \bC$, $ \bc \leq \bchat$ if for all $i=1,\ldots, k_N$, $C_i \subseteq \widehat{C}_j$ for some $j \in \lbrace 1,\ldots, \widehat{k}_N \rbrace$. In this case, the partition space $\bC$ equipped with $\leq$ is a partially ordered set, meaning that the following properties are satisfied:
\begin{itemize}
\item[1.] Reflexivity: $\bc \leq \bc$,
\item[2.] Antisymmetry: if $\bc \leq \bchat$ and $\bchat \leq \bc$ then $ \bc = \bchat$ under a permutation of cluster labels,
\item[3.] Transitivity: if $\bc \leq \bchat$ and $\bchat \leq \widehat{\bchat}$, then $ \bc \leq \widehat{\bchat}$.
\end{itemize}
For any $\bc, \bchat \in \bC$, $\bc$ is covered by $\bchat$, denoted $\bc \prec \bchat$, if $\bc < \bchat$ and there is no $ \widehat{\bchat} \in \bC$ such that $ \bc < \widehat{\bchat} < \bchat$. This covering relation is used to define the \textit{Hasse diagram}, where the elements of $\bC$ are represented as nodes of a graph and a line is drawn from $\bc$ up to $\bchat$ when $\bc \prec \bchat$.

The space of partitions possesses an even richer structure; it forms a lattice. This follows from the fact that every pair of partitions has a \textit{greatest lower bound} and \textit{least upper bound}; for a subset $ \mathbf{S} \subseteq \bC$, an element $\bc \in \bC$ is an upper bound for $ \mathbf{S}$ if $ \mathbf{s}\leq \mathbf{c}$ for all $\mathbf{s} \in \mathbf{S}$, and $\bc \in \bC$ is the least upper bound for $ \mathbf{S}$, denoted $\bc=\text{l.u.b.}(\mathbf{S})$, if $\bc$ is an upper bound for $\mathbf{S}$ and $ \bc \leq \mathbf{c}'$ for all upper bounds $\mathbf{c}'$ of $\mathbf{S}$. A lower bound and the greatest lower bound  for a subset $ \mathbf{S} \subseteq \bC$ are similarly defined, the latter denoted by $\text{g.l.b.}(\mathbf{S})$.
In general, a partially ordered set satisfying these properties forms a lattice; that is, for any $\bc, \bchat, \widehat{\bchat} \in \bC$
\begin{itemize}
\item[1.] $\bc\wedge \bc=\bc$ and $\bc \vee \bc =\bc$,
\item[2.] $\bc \wedge \bchat=\bchat \wedge \bc$ and $\bc \vee \bchat=\bchat \vee \bc$,
\item[3.] $\bc \wedge (\bchat \wedge \widehat{\bchat})=(\bc \wedge \bchat) \wedge \widehat{\bchat}$ and $\bc \vee (\bchat \vee \widehat{\bchat})=(\bc \vee \bchat) \vee \widehat{\bchat}$,
\item[4.] $ \bc \wedge (\bc \vee \bchat)=\bc$ and $\bc \vee (\bc \wedge \bchat)=\bc$,
\end{itemize}
where the operators $\wedge$ and $\vee$ are defined as $\bc \wedge \bchat=\text{g.l.b.}(\bc,\bchat)$ and  $\bc \vee \bchat=\text{l.u.b.}(\bc,\bchat)$ and equality holds under a permutation of cluster labels. In this case, the operator $\wedge$ is called the meet and the operator $\vee$ is called the join. Following the conventions of lattice theory, we will use $ \mathbf{1}$ to denote the greatest element of the lattice of partitions, i.e. the partition with every observation in one cluster $\bc=( \lbrace 1,\ldots,N \rbrace)$, and  $\mathbf{0}$ to denote the least element of the lattice of partitions, i.e. the partition with every observation in its own cluster $\bc= (\lbrace 1\rbrace,\ldots, \lbrace N \rbrace)$. See \cite{Nation91} for more details on lattice theory.

\section{Examples of Balls of Partitions}

\begin{figure}[!h]
\centering
\begin{tikzpicture}[scale=.7]
  \tikzstyle{every node}=[draw=none]
\draw (8,0) node (n1) {\textcolor{gray2}{$\lbrace1,2,3,4\rbrace$}} ;
 \draw (1.25,-1.825) node (n2){\textcolor{gray3}{$\lbrace1 \rbrace \lbrace 2,3,4\rbrace$}};
\draw (3.5,-1.825) node (n3){\textcolor{gray3}{$\lbrace2 \rbrace \lbrace 1,3,4\rbrace$}};
\draw (5.75,-1.825) node (n4){\textcolor{gray3}{$\lbrace3 \rbrace \lbrace 1,2,4\rbrace$}};
\draw (8,-1.825) node (n5){\textcolor{gray3}{$\lbrace4 \rbrace \lbrace 1,2,3\rbrace$}};
\draw (10.25,-2.25) node (n6){$\mathbf{\lbrace 1,2 \rbrace \lbrace 3,4\rbrace}$};
\draw (12.5,-2.25) node (n7){\textcolor{gray5}{$\lbrace1,3 \rbrace \lbrace 2,4\rbrace$}};
\draw (14.75,-2.25) node (n8){\textcolor{gray5}{$\lbrace1,4 \rbrace \lbrace 2,3\rbrace$}};
 \draw (1.25,-3.375) node (n9){\textcolor{gray1}{$\lbrace 1 \rbrace \lbrace 2 \rbrace \lbrace3,4\rbrace$}};
\draw (3.75,-3.375) node (n10){\textcolor{gray4}{$\lbrace 1 \rbrace \lbrace 3 \rbrace \lbrace 2,4\rbrace$}};
\draw (6.25,-3.375) node (n11){\textcolor{gray4}{$\lbrace 1 \rbrace \lbrace 4 \rbrace \lbrace 2,3\rbrace$}};
\draw (8.75,-3.375) node (n12){\textcolor{gray4}{$\lbrace 2 \rbrace \lbrace 3 \rbrace \lbrace 1,4\rbrace$}};
\draw (11.25,-3.375) node (n13){\textcolor{gray4}{$\lbrace 2 \rbrace \lbrace 4 \rbrace \lbrace 1,3\rbrace$}};
\draw (13.75,-3.375) node (n14){\textcolor{gray1}{$\lbrace 3 \rbrace \lbrace 4 \rbrace \lbrace 1,2\rbrace$}};
 \draw (8,-4.5) node (n15){\textcolor{gray2}{$\lbrace 1 \rbrace \lbrace 2 \rbrace \lbrace3 \rbrace \lbrace 4\rbrace$}};

    
  \foreach \from/\to in {n6/n9,n6/n14}
    \draw[gray1] (\from) -- (\to);
    
  \foreach \from/\to in {n1/n6,n9/n15,n14/n15}
    \draw[gray2] (\from) -- (\to);
    
      \foreach \from/\to in {n2/n9,n3/n9,n4/n14,n5/n14}
    \draw[gray3] (\from) -- (\to);
   
      \foreach \from/\to in {n10/n15,n11/n15,n12/n15,n13/n15}
    \draw[gray4] (\from) -- (\to);
    
      \foreach \from/\to in {n7/n10,n7/n13,n8/n11,n8/n12}
    \draw[gray5] (\from) -- (\to);
    
\draw (0,0) -- (0,-4.5);
\draw  (1pt,0 cm) -- (-1pt,0 cm) node[anchor=east] {{\footnotesize $0$}};
\draw  (1pt,-1.825 cm) -- (-1pt,-1.825 cm) node[anchor=east] {{\footnotesize $0.811$}};
\draw  (1pt,-2.25 cm) -- (-1pt,-2.25 cm) node[anchor=east] {{\footnotesize $1$}};
\draw  (1pt,-3.375 cm) -- (-1pt,-3.375 cm) node[anchor=east] {{\footnotesize $1.5$}};
\draw  (1pt,-4.5 cm) -- (-1pt,-4.5 cm) node[anchor=east] {{\footnotesize $2$}};

\end{tikzpicture}
\caption{Example of the VI ball around $\bc=( \lbrace 1,2 \rbrace, \lbrace 3,4 \rbrace)$, with the rainbow color indicating increasing distance from $\bc$. The smallest non-trivial credible ball contains all the red clusterings, the next smallest contains the red and orange clusterings, and so on.}
\label{fig:vi_cb_ex}
\end{figure}
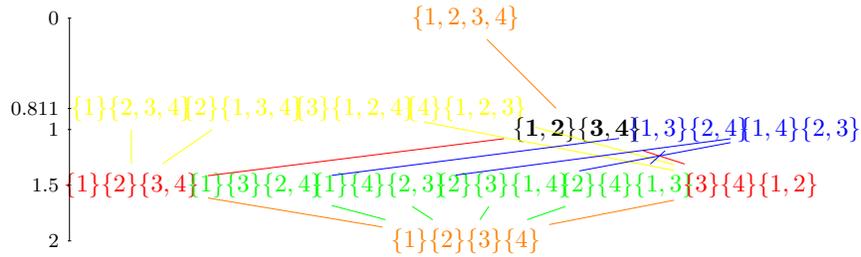

\begin{figure}[!h]
\centering
\begin{tikzpicture}[scale=.7]
  \tikzstyle{every node}=[draw=none]
\draw (8,0) node (n1) {\textcolor{gray4}{$\lbrace1,2,3,4\rbrace$}} ;
 \draw (1.25,-2.25) node (n2){\textcolor{gray3}{$\lbrace1 \rbrace \lbrace 2,3,4\rbrace$}};
\draw (3.5,-2.25) node (n3){\textcolor{gray3}{$\lbrace2 \rbrace \lbrace 1,3,4\rbrace$}};
\draw (5.75,-2.25) node (n4){\textcolor{gray3}{$\lbrace3 \rbrace \lbrace 1,2,4\rbrace$}};
\draw (8,-2.25) node (n5){\textcolor{gray3}{$\lbrace4 \rbrace \lbrace 1,2,3\rbrace$}};
\draw (10.25,-3) node (n6){$\mathbf{\lbrace1,2 \rbrace \lbrace 3,4\rbrace}$};
\draw (12.5,-3) node (n7){\textcolor{gray4}{$\lbrace1,3 \rbrace \lbrace 2,4\rbrace$}};
\draw (14.75,-3) node (n8){\textcolor{gray4}{$\lbrace1,4 \rbrace \lbrace 2,3\rbrace$}};
 \draw (1.25,-3.75) node (n9){\textcolor{gray1}{$\lbrace 1 \rbrace \lbrace 2 \rbrace \lbrace3,4\rbrace$}};
\draw (3.75,-3.75) node (n10){\textcolor{gray3}{$\lbrace 1 \rbrace \lbrace 3 \rbrace \lbrace 2,4\rbrace$}};
\draw (6.25,-3.75) node (n11){\textcolor{gray3}{$\lbrace 1 \rbrace \lbrace 4 \rbrace \lbrace 2,3\rbrace$}};
\draw (8.75,-3.75) node (n12){\textcolor{gray3}{$\lbrace 2 \rbrace \lbrace 3 \rbrace \lbrace 1,4\rbrace$}};
\draw (11.25,-3.75) node (n13){\textcolor{gray3}{$\lbrace 2 \rbrace \lbrace 4 \rbrace \lbrace 1,3\rbrace$}};
\draw (13.75,-3.75) node (n14){\textcolor{gray1}{$\lbrace 3 \rbrace \lbrace 4 \rbrace \lbrace 1,2\rbrace$}};
 \draw (8,-4.5) node (n15){\textcolor{gray2}{$\lbrace 1 \rbrace \lbrace 2 \rbrace \lbrace3 \rbrace \lbrace 4\rbrace$}};

    
  \foreach \from/\to in {n6/n9,n6/n14}
    \draw[gray1] (\from) -- (\to);
    
 \foreach \from/\to in {n9/n15,n14/n15}
   \draw[gray2] (\from) -- (\to);
    
    \foreach \from/\to in {n2/n9,n3/n9,n4/n14,n5/n14,n10/n15,n11/n15,n12/n15,n13/n15}
  \draw[gray3] (\from) -- (\to);

     \foreach \from/\to in {n1/n6,n7/n10,n7/n13,n8/n11,n8/n12}
   \draw[gray4] (\from) -- (\to);
    
\draw (0,0) -- (0,-4.5);
\draw  (1pt,0 cm) -- (-1pt,0 cm) node[anchor=east] {{\footnotesize $0$}};
\draw  (1pt,-2.25 cm) -- (-1pt,-2.25 cm) node[anchor=east] {{\footnotesize $0.375$}};
\draw  (1pt,-3 cm) -- (-1pt,-3 cm) node[anchor=east] {{\footnotesize $0.5$}};
\draw  (1pt,-3.75 cm) -- (-1pt,-3.75 cm) node[anchor=east] {{\footnotesize $0.625$}};
\draw  (1pt,-4.5 cm) -- (-1pt,-4.5 cm) node[anchor=east] {{\footnotesize $0.75$}};

\end{tikzpicture}
\caption{Example of the $\tB$ ball around $\bc=( \lbrace 1,2 \rbrace, \lbrace 3,4 \rbrace)$, with the rainbow color indicating increasing distance from $\bc$. The smallest non-trivial credible ball contains all the red clusterings, the next smallest contains the red and orange clusterings, and so on.}
\label{fig:bl_cb_ex}
\end{figure}

As both VI and $\tB$ are metrics on the space of partitions, we can construct a ball around a partition $\bc$ of size $\epsilon$, defined as:
$$ B_\epsilon(\bc)=\lbrace \bchat \in \bC : d(\bc,\bchat) \leq \epsilon \rbrace.$$
Interestingly, these balls differ between the two metrics even for the simple example with $N=4$ in Figures \ref{fig:vi_cb_ex} and \ref{fig:bl_cb_ex}, which consider a ball around $\bc=( \lbrace 1,2 \rbrace, \lbrace 3,4 \rbrace)$. In these figures, partitions are rainbow colored by increasing distance to $\bc$; thus, the smallest non-trivial ball, i.e. the smallest ball around $\bc$ with at least two partitions, contains all red clusterings, the next smallest ball contains all red and orange clusterings, and so on. 
In general, from Property \ref{prop:closest_part}, the smallest non-trivial ball will be the same for the two metrics, which is confirmed in the figures, as the set of red clusterings coincide. When considering the next smallest ball, differences emerge; the VI ball includes the red clusterings and the orange clusterings $\0$ and $\1$, and the $\tB$ ball includes the red clusterings and the orange clustering $\0$. Note that $\1$ is only included in the $\tB$ ball around $\bc$ when it is expanded to include all clusterings and is considered as distant to $\bc$ as the partitions $\bc=( \lbrace 1,3 \rbrace, \lbrace 2,4 \rbrace)$ and $\bc=( \lbrace 1,4 \rbrace, \lbrace 2,3 \rbrace)$.  In the authors' opinions, the VI ball more closely reflects our intuition of the closest set of partitions to $\bc$.

Further suppose that $N=4$ and the optimal point estimate under both VI and $\tB$ is $(\lbrace 1,2 \rbrace, \lbrace 3,4 \rbrace)$. Say that the $95\%$ credible ball with the VI and $\tB$ metric consists of the red, orange, and yellow partitions depicted in Figures \ref{fig:vi_cb_ex} and \ref{fig:bl_cb_ex}, respectively. For VI, the upper vertical bound is $\1$, the lower vertical bound is $\0$, and the horizontal bounds are the yellow partitions, i.e. those with one singleton and one cluster of size $N-1=3$. 
For $\tB$, the upper vertical bounds are the partitions with one singleton and one cluster of size $N-1=3$, the lower vertical bound is $\0$, and the horizontal bounds are the yellow partitions, i.e. those with one singleton and one cluster of size $N-1=3$ and those with two singletons and one cluster of size $N-2=2$ which are not covered by $\bc^*$.

\section{Properties and Proofs}

\begin{property}
Both VI and $\tB$ are metrics on the space of partitions, that is they satisfy:
\begin{itemize}
\item[1.] Non-negativity: $d(\bc,\bchat) \geq 0$  and $d(\bc,\bchat)=0$ if and only if $ \bc=\bchat$ under a permutation of cluster labels,
\item[2.] Symmetry: $d(\bc,\bchat) =d(\bchat,\bc)$,
\item[3.] Triangle inequality: for any $\bc,\bchat, \widehat{\bchat}$,
$$d(\bc,\bchat)\leq d(\bc,\widehat{\bchat})+d(\bchat, \widehat{\bchat}).$$
\end{itemize}
\end{property}
A proof for VI can be found in \cite{Meila07}. For $\tB$, the proof results from the fact that $\tB$ can be derived as the Hamming distance between the binary representation of the clusterings.

\begin{property} \label{prop:VertAlign}
For both VI and $\tB$, if $ \bc \geq \bchat \geq \widehat{\bchat}$, 
then
$$ d(\bc, \widehat{\bchat})= d(\bc, \bchat)+ d(\bchat,\widehat{\bchat}).$$
\end{property}

\noindent{\em Proof of Property \ref{prop:VertAlign}}.
First note that if $\bc \geq \bchat$, or equivalently $\bc \wedge \bchat =\bchat$, then for all nonzero $n_{i\,j}$, $n_{i\,j}=n_{+\,j}$. Thus, if $\bc \geq \bchat$,
\begin{align}
\VI(\bc,\bchat)&=\sum_{i=1}^{k_N} \frac{n_{i\,+}}{N}\log\left(\frac{n_{i\,+}}{N}\right)+\sum_{j=1}^{\widehat{k}_N} \frac{n_{+\,j}}{N}\log\left(\frac{n_{+\,j}}{N}\right)-2\sum_{j=1}^{\widehat{k}_N} \frac{n_{+\,j}}{N}\log\left(\frac{n_{+\,j}}{N}\right) \nonumber \\
&=\sum_{i=1}^{k_N} \frac{n_{i\,+}}{N}\log\left(\frac{n_{i\,+}}{N}\right)-\sum_{j=1}^{\widehat{k}_N} \frac{n_{+\,j}}{N}\log\left(\frac{n_{+\,j}}{N}\right). \label{eq:VIvert}
\end{align}
Let $n_i$ denote the number of observations in cluster $i$ under $\bc$; $ \widehat{n}_i$ denote the number of observations in cluster $i$ under $\bchat$; and $\widehat{\widehat{n}}_i$ denote the number of observations in cluster $i$ under $\widehat{\bchat}$. Then, from (\ref{eq:VIvert}), if $\bc \geq \bchat \geq \widehat{\bchat}$,
\begin{align*}
\VI(\bc,\widehat{\bchat})&=\sum_{i=1}^{k_N} \frac{n_{i}}{N}\log\left(\frac{n_{i}}{N}\right)-\sum_{i=1}^{\widehat{\widehat{k}}_N} \frac{\widehat{\widehat{n}}_{i}}{N}\log\left(\frac{\widehat{\widehat{n}}_{i}}{N}\right)\\
&=\sum_{i=1}^{k_N} \frac{n_{i}}{N}\log\left(\frac{n_{i}}{N}\right)-\sum_{i=1}^{\widehat{k}_N} \frac{\widehat{n}_{i}}{N}\log\left(\frac{\widehat{n}_{i}}{N}\right)+\sum_{i=1}^{\widehat{k}_N} \frac{\widehat{n}_{i}}{N}\log\left(\frac{\widehat{n}_{i}}{N}\right)-\sum_{i=1}^{\widehat{\widehat{k}}_N} \frac{\widehat{\widehat{n}}_{i}}{N}\log\left(\frac{\widehat{\widehat{n}}_{i}}{N}\right)\\
&=\VI(\bc,\bchat)+\VI(\bchat,\widehat{\bchat})
\end{align*}
For $\tB$ the proof is similar, since if $\bc \geq \bchat$,
\begin{align*}
\tB(\bc,\bchat)&=\sum_{i=1}^{k_N} \left( \frac{n_{i\,+}}{N} \right)^2-\sum_{j=1}^{\widehat{k}_N} \left(\frac{n_{+\,j}}{N}\right)^2. 
\end{align*} 
\qed
\\

\begin{property} \label{prop:HorzAlign}
For both VI and $\tB$,
$$ d(\bc, \bchat)= d(\bc, \bchat\wedge \bc)+ d(\bchat,\bchat\wedge \bc).$$
\end{property}
\noindent{\em Proof of Property \ref{prop:HorzAlign}}.
The meet between two clusterings $\bc$ and $\bchat$ will have at most $k_N\cdot\widehat{k}_N$ clusters with $n_{i\,j}$ data points in cluster $i,j$ for $i=1,\ldots,k_N$, $j=1,\ldots, \widehat{k}_N$ (some $n_{i\,j}$ may equal zero, resulting in less than $k_N\cdot\widehat{k}_N$ clusters).
\begin{align*}
\VI(\bc,\bchat)&=\sum_{i=1}^{k_N} \frac{n_{i\,+}}{N}\log\left(\frac{n_{i\,+}}{N}\right)+\sum_{j=1}^{\widehat{k}_N} \frac{n_{+\,j}}{N}\log\left(\frac{n_{+\,j}}{N}\right)-2\sum_{i=1}^{k_N}\sum_{j=1}^{\widehat{k}_N} \frac{n_{i\,j}}{N}\log\left(\frac{n_{i\,j}}{N}\right) \\
&=\sum_{i=1}^{k_N} \frac{n_{i\,+}}{N}\log\left(\frac{n_{i\,+}}{N}\right)-\sum_{i=1}^{k_N}\sum_{j=1}^{\widehat{k}_N} \frac{n_{i\,j}}{N}\log\left(\frac{n_{i\,j}}{N}\right)+\sum_{j=1}^{\widehat{k}_N} \frac{n_{+\,j}}{N}\log\left(\frac{n_{+\,j}}{N}\right)-\sum_{i=1}^{k_N}\sum_{j=1}^{\widehat{k}_N} \frac{n_{i\,j}}{N}\log\left(\frac{n_{i\,j}}{N}\right)\\
&=\VI(\bc,\bc \wedge \bchat)+\VI(\bchat,\bc\wedge\bchat).
\end{align*}
The last line follows from the fact that the pairs $(\bc,\bc\wedge\bchat)$ and $(\bchat,\bc\wedge\bchat)$ are vertically aligned, see the proof of Property \ref{prop:VertAlign}. The proof for $\tB$ is similar.
\qed
\\

\begin{property} \label{prop:scale}
A distance on partitions satisfying Properties \ref{prop:VertAlign} and \ref{prop:HorzAlign} has the property that for any two partitions $\bc$ and $\bchat$,
\begin{align*}
d(\bc, \bchat) \leq d(\1,\0).
\end{align*}
Thus,
$$ \VI(\bc, \bchat) \leq \log(N)\quad \text{  and  } \quad \tB(\bc, \bchat) \leq 1-\frac{1}{N}.$$
\end{property}
\noindent{\em Proof of Property \ref{prop:scale}}.
First note that from Property \ref{prop:VertAlign}
\begin{align}
d(\1,\0)=d(\1,\bc)+d(\bc,\bc\wedge\bchat)+d(\0,\bc \wedge \bchat), \label{eq:d10_rep1}
\end{align}
and
\begin{align}
d(\1,\0)=d(\1,\bchat)+d(\bchat,\bc\wedge\bchat)+d(\0,\bc \wedge \bchat). \label{eq:d10_rep2}
\end{align}
Combining (\ref{eq:d10_rep1}) and (\ref{eq:d10_rep2}), we have
\begin{align*}
2d(\1,\0)=d(\1,\bc)+d(\1,\bchat)+d(\bc,\bc\wedge\bchat)+d(\bchat,\bc\wedge\bchat)+2d(\0,\bc \wedge \bchat),
\end{align*}
which, using Property \ref{prop:HorzAlign}, implies that
\begin{align*}
d(\1,\0)=\frac{1}{2} \left(d(\1,\bc)+d(\1,\bchat)+d(\bc,\bchat)\right)+d(\0,\bc \wedge \bchat).
\end{align*}
Since $d$ is a metric, the triangle inequality states that
\begin{align*}
d(\bc,\bchat) \leq d(\1,\bc)+d(\1,\bchat).
\end{align*}
Adding $d(\bc,\bchat)$ to either side, we have
\begin{align*}
d(\bc,\bchat) \leq\frac{1}{2} \left(d(\1,\bc)+d(\1,\bchat) +d(\bc,\bchat) \right).
\end{align*}
Thus,
\begin{align*}
d(\bc,\bchat) \leq&\frac{1}{2} \left(d(\1,\bc)+d(\1,\bchat) +d(\bc,\bchat) \right) +d(\0,\bc \wedge \bchat)-d(\0,\bc \wedge \bchat)\\
&= d(\1,\0)-d(\0,\bc \wedge \bchat) \leq d(\1,\0).
\end{align*}
The last two statements results from $\VI(\1,\0)=\log(N)$ and $\tB(\1,\0)=1-\frac{1}{N}$.
\qed
\\

\begin{property}\label{prop:closest_part}
For both metrics $\VI$ and $\tB$, the closest partitions to a partition $\bc$ are:
\begin{itemize}
\item if $\bc$ contains at least two clusters of size one and at least one cluster of size two, the partitions which merge any two clusters of size one and the partitions which split any cluster of size two.
\item if $\bc$ contains at least two clusters of size one and no clusters of size two, the partitions which merge any two clusters of size one.
\item if $\bc$ contains at most one cluster of size one, the partitions which split the smallest cluster of size greater than one into a singleton and a cluster with the remaining observations of the original cluster. 
\end{itemize}
\end{property}
\noindent{\em Proof of \ref{prop:closest_part}}.
From Property \ref{prop:HorzAlign}, the distance between $\bc$ and any $\bchat \neq \bc$ will be bounded below by the distance between $\bc$ and their meet $\bc \wedge \bchat$,
$$ \VI(\bc, \bchat) \geq \VI(\bc, \bc \wedge \bchat).$$
If $\bc \wedge \bchat =\bc$, then $\bchat > \bc$, and there exists a $\bc^m$ such that $\bchat \geq \bc^m \succ \bc$. From Property \ref{prop:VertAlign}, $\VI(\bc,\bchat) \geq \VI(\bc,\bc^m)$. Otherwise, $\bc \wedge \bchat <\bc$, and there exists a $\bc^s$ such that $ \bc \wedge \bchat \leq \bc^s \prec \bc$. From Property \ref{prop:VertAlign}, $\VI(\bc,\bc \wedge \bchat) \geq \VI(\bc,\bc^s)$. Thus the closest partitions to $\bc$ will be among those which cover $\bc$ or those which $\bc$ covers.

Let's first consider the partitions which cover $\bc$. If $\bc^m \succ \bc$, then $\bc^m$ is obtained from $\bc$ by merging two clusters $i$ and $j$, and
\begin{align}
\VI(\bc,\bc^m)= \frac{1}{N} \left( (n_i+n_j) \log(n_i+n_j)- n_{i}\log(n_{i})-n_{j}\log(n_{j}) \right), \label{eq:merge}
\end{align}
which is minimized when $i$ and $j$ correspond to the two smallest clusters in $\bc$. Next consider the partitions which $\bc$ covers. If $\bc^s \prec \bc$, then $\bc^s$ is obtained from $\bc$ by splitting a cluster $i$ of size $n_i>1$ into two clusters $i_1, i_2$ of sizes $n_{i_1}$ and $n_{i_2}$, and 
\begin{align}
\VI(\bc,\bc^s)= \frac{1}{N} \left( n_i \log(n_i)- n_{i_1}\log(n_{i_1})-n_{i_2}\log(n_{i_2}) \right), \label{eq:split}
\end{align}
which is minimized when $i$ is the smallest cluster of size $n_i>1$ and $n_{i_1}=1$ and $n_{i_2}=n_i-1$. Thus, the closest partitions to $\bc$ will be among those which merge the two smallest clusters or split the smallest cluster $i$ of size $n_i>1$ into a singleton and a cluster of size $n_i-1$.

To compare (\ref{eq:merge}) and (\ref{eq:split}), we first consider the case when $\bc$ contains at least two singletons. In this case, the closest partitions which cover $\bc$ merge two singletons, and (\ref{eq:merge}) reduces to
$$\frac{1}{N} \left( 2\log(2)- 1\log(1)-1\log(1) \right)=\frac{2}{N}.$$
Letting $i$ be the index of the smallest cluster of size $n_i>1$, the closest partition which $\bc$ covers  splits cluster $i$ into a singleton and a cluster of size $n_i-1$, and (\ref{eq:split}) reduces to
$$\frac{1}{N} \left( n_i \log(n_i)-(n_{i}-1)\log(n_{i}-1) \right).$$
For $n_i=2$, these distances are equal, and the closest partitions to $\bc$ will be those which merge any two singletons or split any cluster of size $n_i=2$. For $n_i>2$, the partitions which merge any two singletons are the closest. 

Next, suppose that $\bc$ contains at most one singleton, and let $n_i,n_j$ denote the sizes of the smallest two clusters, where $n_i$ corresponds to the size of the smaller cluster unless a singleton is present. The closest partitions which cover $\bc$ merge any two clusters of sizes $n_i$ and $n_j$, and (\ref{eq:merge}) is
\begin{align}
\frac{1}{N} \left( (n_i+n_j)\log(n_i+n_j)- n_i\log(n_i)-n_j\log(n_j) \right). \label{eq:merge_sc}
\end{align}
The closest partitions which $\bc$ covers split any cluster of size $n_i$ into a singleton and a cluster of size $n_i-1$, (\ref{eq:split}) reduces to
\begin{align}
\frac{1}{N} \left( n_i \log(n_i)-(n_{i}-1)\log(n_{i}-1) \right). \label{eq:split_sc}
\end{align}
In this case, (\ref{eq:split_sc}) will be less than (\ref{eq:merge_sc}), and the closest partitions are those which split the smallest cluster $i$ into a singleton and cluster of size $n_i-1$. The proof for $\tB$ is similar.
\qed
\\

\begin{property} \label{prop:symmetry}
Suppose $N$ is divisible by $k$, and let $\bc_k$ denote a partition with $k$ clusters of equal size $N/k$.
$$ \tB(\1, \bc_k)= 1-\frac{1}{k} > \frac{1}{k}-\frac{1}{N}= \tB(\0,\bc_k).$$
$$ \VI(\1, \bc_k)= \log(k)  \leq \log(N)-\log(k) =\VI(\0, \bc_k), \quad \text{for } k \leq \sqrt{N},$$
and
$$ \VI(\1, \bc_k)= \log(k)  \geq \log(N)-\log(k) =\VI(\0, \bc_k), \quad \text{for } k \geq \sqrt{N}.$$
\end{property}

\begin{property}
Suppose $N$ is an even and square integer. Then, the partitions with two clusters of sizes $n=\frac{1}{2}(N-\sqrt{N})$ and $N-n$ are equally distant from $\1$ and $\0$ under $\tB$.
\end{property}

\section*{Appendix III: Examples}

\begin{table}[!h]
\centering
\subfloat[{\normalsize \textbf{Binder's: 9 clusters}}]{\begin{tabular}{c|ccccccccc}
$\bc_t \setminus \bc^*$ & 1  &2  &3  &4  &5  &6  &7 & 8  &9\\ \hline
                  1 &\textbf{41} & 1  &0  &0  &0  &0  &1  &0  &1\\
                  2  &1 &\textbf{52} & 0&  0&  1&  0&  0&  0&  0\\
                  3 & 1 & 0 & \textbf{44} & 3 & 1 & 0 & 0 & 0&  0\\
                  4&  0 & 1 & 0 & 0&\textbf{ 50} & 1 & 0 & 1 & 0
\end{tabular}\label{tbl:crosstab_ex1B}} \\
\subfloat[{\normalsize \textbf{VI: 4 clusters}}]{\begin{tabular}{c|cccc}
$\bc_t \setminus \bc^*$ & 1 & 2 & 3 & 4\\ \hline
                  1 &\textbf{42} & 2 & 0 & 0\\
                  2 & 1 &\textbf{52} & 0 & 1\\
                  3  &1 & 0 &\textbf{45} & 3\\
                  4  &0 & 1 & 0 &\textbf{52}\\
\end{tabular}\label{tbl:crosstab_ex1VI}}
\caption{\textbf{Example 1}: cross tabulation of the true cluster labels and \textbf{estimated} cluster labels.}
\label{tbl:crosstab_ex1}
\end{table}

\begin{table}[!h]
\centering
\subfloat[{\normalsize \textbf{Binder's: 12 clusters}}]{\begin{tabular}{c|cccccccccccc}
$\bc_t \setminus \bc^*$ & 1  &2  &3  &4  &5  &6  &7 & 8  &9 & 10 & 11 & 12\\ \hline
                  1 &\textbf{44} & 0 & 0 & 0 & 0 & 0 & 0 & 0 & 0 & 0 & 0 & 0\\
                  2 & 0 & 0 & 0 & 0 & 0 & 1 & 0 &\textbf{53} & 0 & 0 & 0 & 0\\
                  3 & 0 & 0 & 0 &\textbf{45} & 2 & 0 & 0 & 0 & 1 & 1 & 0 & 0\\
                  4 & 0 & 1 & 1 & 2 & 2 &\textbf{40} & 2 & 3 & 0 & 0 & 1 & 1
\end{tabular}\label{tbl:crosstab_ex2B}} \\
\subfloat[{\normalsize \textbf{VI: 4 clusters}}]{\begin{tabular}{c|cccc}
$\bc_t \setminus \bc^*$ & 1 & 2 & 3 & 4\\ \hline
                  1 &\textbf{44} & 0 & 0 & 0\\
                  2 & 0 &\textbf{53} & 0 & 1\\
                  3 & 0 & 0 &\textbf{47} & 2\\
                  4 & 0 & 4 & 3& \textbf{46}
\end{tabular}\label{tbl:crosstab_ex2VI}}
\caption{\textbf{Example 2}: cross tabulation of the true cluster labels and \textbf{estimated} cluster labels. The first cluster with decreased variance is well identified in both estimates, as is clearly evident from the first row and column. }
\label{tbl:crosstab_ex2}
\end{table}

%

\begin{table}[!h]
\centering
\subfloat[{\normalsize \textbf{Upper vertical bound: 4 clusters}}]{\begin{tabular}{c|cccc}
$\bc_t \setminus \bc_{u}$ & 1  &2  &3  &4  \\ \hline
                  1 &43& 1  &0  &0  \\
                  2  &0 &52& 0&  2\\
                  3 & 1 & 0 & 29 & 19 \\
                  4&  1 & 2  & 0&50
\end{tabular}\label{tbl:crosstab_ex1B_upper}}\\
\subfloat[{\normalsize \textbf{Lower vertical bound: 18 clusters}}]{\begin{tabular}{c|cccccccccccccccccc}
$\bc_t \setminus \bc_{l}$ & 1  &2  &3  &4 & 5 &6 & 7 & 8 & 9 &10 &11 &12 &13 &14 &15 &16 &17 &18  \\ \hline
                  1 &39 & 3 & 0 & 0 & 0 & 0 & 0 & 0 & 1 & 1&  0&  0&  0 & 0 & 0 & 0 & 0&  0\\
                  2 & 0 &43&  0&  0&  0&  0&  0 & 1 & 1 & 3 & 0 & 5&  1&  0&  0&  0 & 0&  0\\
                  3 & 2 & 0 &40&  1&  2&  0&  0&  0&  0 & 0&  0&  0&  0 & 2 & 1 & 0 & 0&  1\\
                  4 & 0&  1&  0&  0& 23 & 1 & 1 &20 & 0 & 0 & 1 & 2&  0&  0&  2&  1&  1&  0

\end{tabular}\label{tbl:crosstab_ex1B_lower}}\\
\subfloat[{\normalsize \textbf{Horizontal bound: 11 clusters}}]{\begin{tabular}{c|ccccccccccc}
$\bc_t \setminus \bc_{h}$ & 1  &2  &3  &4  & 5 & 6 & 7 & 8 & 9 &10 &11\\ \hline
                  1 &41&  1&  0&  0 & 0 & 1 & 0 & 0 & 1 & 0 & 0\\
                  2 & 0 &36&  0&  1&  0& 11&  2&  0 & 4 & 0 & 0\\
                  3 & 1&  0& 43&  0&  2&  1&  0&  1&  0&  0&  1\\
                  4 & 0 & 0 & 3&  4& 42&  0&  3&  0 & 0 & 1 & 0
\end{tabular}\label{tbl:crosstab_ex1B_horiz}}\\
\caption{\textbf{Example 1}: cross tabulation of the true cluster labels and \textbf{credible bound} cluster labels from \textbf{Binder's}.}
\label{tbl:crosstab_ex1B_cb}
\end{table}

\begin{table}[!h]
\centering
\subfloat[{\normalsize \textbf{Upper vertical bound: 4 clusters}}]{\begin{tabular}{c|cccc}
$\bc_t \setminus \bc_{u}$ & 1  &2  &3  &4  \\ \hline
                  1 &40&  2&  2&  0\\
                  2 & 0 &52 & 0 & 2\\
                  3 & 1 & 0 &43 & 5\\
                  4 & 0 & 6 & 4 &43
\end{tabular}\label{tbl:crosstab_ex1VI_upper}}\\
\subfloat[{\normalsize \textbf{Lower vertical bound: 16 clusters}}]{\begin{tabular}{c|cccccccccccccccc}
$\bc_t \setminus \bc_{l}$ & 1  &2  &3  &4 & 5 &6 & 7 & 8 & 9 &10 &11 &12 &13 &14 &15 &16   \\ \hline
                  1 &27 & 3 &12 & 0 & 0 & 0 & 0  &0  &1 & 0&  1&  0 & 0 & 0 & 0 & 0\\
                  2 & 1 &43 & 0&  0 & 0 & 1 & 4 & 0 & 1 & 0&  0&  3 & 1 & 0 & 0 & 0\\
                  3 & 0&  0 & 0 &44 & 1 & 1 & 0 & 0 & 0 & 1 & 0 & 0 & 0 & 1 & 0 & 1\\
                  4&  0&  0 & 0 & 0 & 0 &45 & 1 & 6 & 0 & 0 & 0 & 0 & 0 & 0 & 1 & 0
\end{tabular}\label{tbl:crosstab_ex1VI_lower}}\\
\subfloat[{\normalsize \textbf{Horizontal bound: 11 clusters}}]{\begin{tabular}{c|ccccccccccc}
$\bc_t \setminus \bc_{h}$ & 1  &2  &3  &4  & 5 & 6 & 7 & 8 & 9 &10 &11\\ \hline
                  1 &35 & 1 & 2 & 0 & 0 & 0 & 4 & 0 & 1 & 0 & 1\\
                  2 & 1 &50 & 0 & 0 & 0 & 1 & 0  &0 & 1 & 1 & 0\\
                  3 & 0 & 0 & 0 &15 &32 & 1 & 1& 0 & 0 & 0 & 0\\
                  4  &0 & 2 & 0& 0 & 4 &21 & 1 &25 & 0 & 0 & 0
\end{tabular}\label{tbl:crosstab_ex1VI_horiz}}\\
\caption{\textbf{Example 1}: cross tabulation of the true cluster labels and \textbf{credible bound} cluster labels from \textbf{VI}.}
\label{tbl:crosstab_ex1VI_cb}
\end{table}

\begin{table}[!h]
\centering
\subfloat[{\normalsize \textbf{Upper vertical bounds: 4 clusters}}]{\begin{tabular}{c|cccc}
$\bc_t \setminus \bc_{u}$ & 1  &2  &3  &4  \\ \hline
                  1 &44 & 0 & 0 & 0\\
                  2 & 0 &25 &29 & 0\\
                  3 & 0 & 3 & 0 &46\\
                  4 & 0 &49 & 1 & 3
\end{tabular}\label{tbl:crosstab_ex2B_upper1}} \\ 
\subfloat[{\normalsize \textbf{Lower vertical bound: 19 clusters}}]{\begin{tabular}{c|ccccccccccccccccccc}
$\bc_t \setminus \bc_{l}$ & 1  &2  &3  &4 & 5 &6 & 7 & 8 & 9 &10 &11 &12 &13 &14 &15 &16 &17 &18& 19  \\ \hline
                  1 &44 & 0 & 0 & 0 & 0 & 0 & 0 & 0 & 0 & 0 & 0 & 0 & 0 & 0 & 0 & 0 & 0&  0&  0\\
                  2 & 0& 34&  0&  0& 10 & 1 & 1 & 4 & 0 & 3 & 0 & 0 & 1 & 0 & 0 & 0 & 0 & 0&  0\\
                  3 & 0 & 0 &39 & 0 & 0 & 0 & 0 & 0 & 0 & 0 & 0 & 2 & 0 & 6 & 0 & 1 & 0 & 0 & 1\\
                  4 & 0 & 1 &17 & 1&  1& 14&  0&  3& 11 & 0 & 1 & 0 & 0 & 0 & 1 & 0 & 2&  1 & 0
\end{tabular}\label{tbl:crosstab_ex2B_lower}}\\
\subfloat[{\normalsize \textbf{Horizontal bound: 10 clusters}}]{\begin{tabular}{c|cccccccccc}
$\bc_t \setminus \bc_{h}$ & 1  &2  &3  &4  & 5 & 6 & 7 & 8 & 9 &10 \\ \hline
                  1 &44 & 0 & 0 & 0 & 0 & 0 & 0 & 0 & 0 & 0\\
                  2  &0 &42  &0  &2 & 2 & 3  &2  &1  &2 & 0\\
                  3  &0  &1 &24& 23&  0 & 0 & 0 & 0 & 1 & 0\\
                  4 & 0 & 2 & 1 &30 & 3 &14 & 0 & 0 & 0 & 3
\end{tabular}\label{tbl:crosstab_ex2B_horiz}}\\
\caption{\textbf{Example 2}: cross tabulation of the true cluster labels and \textbf{credible bound} cluster labels from \textbf{Binder's}.}
\label{tbl:crosstab_ex2B_cb}
\end{table}

\begin{table}[!h]
\centering
\subfloat[{\normalsize \textbf{Upper vertical bound: 3 clusters}}]{\begin{tabular}{c|ccc}
$\bc_t \setminus \bc_{u}$ & 1  &2  &3   \\ \hline
                  1 &44 & 0 & 0\\
                  2 & 0 &53 & 1\\
                  3 & 1 & 1 &47\\
                  4  &0 &36& 17
\end{tabular}\label{tbl:crosstab_ex2VI_upper}}\\
\subfloat[{\normalsize \textbf{Lower vertical bound: 16 clusters}}]{\begin{tabular}{c|cccccccccccccccc}
$\bc_t \setminus \bc_{l}$ & 1  &2  &3  &4 & 5 &6 & 7 & 8 & 9 &10 &11 &12 &13 &14 &15 &16   \\ \hline
                  1 &43 & 0 & 0 & 0 & 0 & 0 & 0&  0 & 0 & 0& 0 & 0 & 1 & 0 & 0 & 0\\
                  2  &0 &42 & 0 & 0 & 0 & 1 & 1  &3 & 4 & 0 & 0 & 0 & 1 & 0 & 1 & 1\\
                  3 & 0 & 0 &41 & 0 & 0 & 0 & 0 & 0 & 0 & 4 & 1 & 1 & 0 & 2 & 0 & 0\\
                  4  &0 & 0 & 3 & 2 & 1 &27 & 2 & 4 &13 & 0 & 0 & 0 & 0 & 0 & 1 & 0
\end{tabular}\label{tbl:crosstab_ex2VI_lower}}\\
\subfloat[{\normalsize \textbf{Horizontal bound: 6 clusters}}]{\begin{tabular}{c|cccccc}
$\bc_t \setminus \bc_{h}$ & 1  &2  &3  &4  & 5 & 6 \\ \hline
                  1 &43 & 1 & 0 & 0 & 0 & 0\\
                  2 & 0 &53 & 0 & 0 & 1 & 0\\
                  3  &0  &1& 26& 21 & 0 & 1\\
                  4 & 0 &11 & 2 &28 &12 & 0
\end{tabular}\label{tbl:crosstab_ex2VI_horiz}}\\
\caption{\textbf{Example 2}: cross tabulation of the true cluster labels and \textbf{credible bound} cluster labels from \textbf{VI}.}
\label{tbl:crosstab_ex2VI_cb}
\end{table}

\begin{table}[!h]
\centering
\subfloat[{\normalsize \textbf{Upper vertical bound: 2 clusters}}]{\begin{tabular}{c|cccc}
$\bc^* \setminus \bc_{u}$ & 1  &2  \\ \hline
        1 & 7 & 0\\
        2 &30 &42\\
        3  &3 & 0
\end{tabular}\label{tbl:crosstab_galaxy_upper}}\\
\subfloat[{\normalsize \textbf{Lower vertical bound: 15 clusters}}]{\begin{tabular}{c|cccccccccccccccccc}
$\bc^* \setminus \bc_{l}$ & 1  &2  &3  &4 & 5 &6 & 7 & 8 & 9 &10 &11 &12 &13 &14 &15   \\ \hline
     1  &2 & 1 & 3 & 1 & 0 & 0 & 0 & 0 & 0 & 0 & 0 & 0 & 0 & 0 & 0\\
        2 & 2 & 0 & 0 & 0 & 1 &11 & 1& 50 & 3 & 1 & 1 & 1 & 1 & 0 & 0\\
        3 & 0 & 0 & 0 & 0 & 0 & 0 & 0 & 0 & 0&  0 & 0 & 0 & 0 & 2 & 1
\end{tabular}\label{tbl:crosstab_galaxy_lower}}\\
\subfloat[{\normalsize \textbf{Horizontal bound: 8 clusters}}]{\begin{tabular}{c|ccccccccccc}
$\bc^* \setminus \bc_{h}$ & 1  &2  &3  &4  & 5 & 6 & 7 & 8 \\ \hline
        1  &7 & 0 & 0 & 0 & 0 & 0 & 0 & 0\\
        2 & 6 & 1& 31 & 1& 28 & 4 & 1 & 0\\
        3  &0 & 0&  0&  0 & 0 & 0 & 1 & 2
\end{tabular}\label{tbl:crosstab_galaxy_horiz}}\\
\caption{\textbf{Galaxy example}: cross tabulation of the estimated cluster labels and credible bound cluster labels from VI.}
\label{tbl:crosstab_galaxy_cb}
\end{table}

 \bibliographystyle{ba}
\bibliography{main}